\documentclass[3p]{elsarticle}

\usepackage{lineno,hyperref}

\usepackage[T1]{fontenc}
\usepackage[utf8]{inputenc}
\usepackage{mathtools}
\usepackage{amsfonts}
\usepackage{amssymb}
\usepackage{esint}
\usepackage{graphicx}
\usepackage{mathrsfs}
\usepackage{dsfont}
\usepackage{booktabs}
\usepackage{multirow}
\usepackage[nameinlink,capitalize]{cleveref}
\usepackage{siunitx}
\usepackage{float}
\usepackage{xstring}
\usepackage{subcaption}
\usepackage{todonotes}

\crefname{appendix}{}{}

\graphicspath{{Figures/}}

\renewcommand{\d}{\mathrm{d}}
\newcommand{\ga}{g_{A}}
\newcommand{\fpi}{f_{\pi}}
\newcommand{\mpi}{m_{\pi}}
\newcommand{\mnuc}{M_{N}}

\renewcommand{\Im}{\operatorname{Im}} 
\renewcommand{\d}{\mathrm{d}}
\newcommand{\e}{\mathrm{e}}
\newcommand{\iu}{\mathrm{i}}
\newcommand{\1}{\ensuremath{\mathds{1}}}
\newcommand{\melement}[3]{\left<#1\left|\vphantom{#1}#2\vphantom{#3}\right|#3\right>}

\newcommand{\ket}[1]{\left|#1\right>}
\DeclarePairedDelimiter{\abs}{\lvert}{\rvert}
\DeclareRobustCommand{\pwave}[1]{\ensuremath{{}^{\StrChar{#1}{1}}{\StrChar{#1}{2}}_{\StrChar{#1}{3}}}}

\newcommand{\diagrambox}[1][\defaultscale]{\raisebox{-.35\height}{\includegraphics[scale=#1]{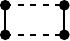}}}
\newcommand{\diagramboxcrossed}[1][\defaultscale]{\raisebox{-.35\height}{\includegraphics[scale=#1]{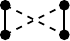}}}

\newcommand{\diagramboxDD}[1][\defaultscale]{\raisebox{-.35\height}{\includegraphics[scale=#1]{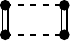}}}
\newcommand{\diagramboxcrossedDD}[1][\defaultscale]{\raisebox{-.35\height}{\includegraphics[scale=#1]{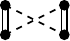}}}
\newcommand{\diagramtriangle}[1][\defaultscale]{\raisebox{-.35\height}{\includegraphics[scale=#1]{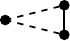}}}
\newcommand{\diagramtriangleD}[1][\defaultscale]{\raisebox{-.35\height}{\includegraphics[scale=#1]{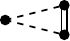}}}
\newcommand{\diagramfootball}[1][\defaultscale]{\raisebox{-.35\height}{\includegraphics[scale=#1]{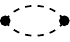}}}
\newcommand{\diagramboxDmirrored}[1][\defaultscale]{\raisebox{-.35\height}{\reflectbox{\includegraphics[scale=#1]{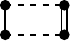}}}}
\newcommand{\diagramboxcrossedDmirrored}[1][\defaultscale]{\raisebox{-.35\height}{\reflectbox{\includegraphics[scale=#1]{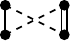}}}}
\newcommand{\diagramtrianglemirrored}[1][\defaultscale]{\raisebox{-.35\height}{\reflectbox{\includegraphics[scale=#1]{Diagram_Triangle.pdf}}}}
\newcommand{\diagramtriangleDmirrored}[1][\defaultscale]{	\raisebox{-.35\height}{\reflectbox{\includegraphics[scale=#1]{Diagram_TriangleD.pdf}}}}
\newcommand{\propagatorN}[1][1.0]{	\raisebox{-.35\height}{\includegraphics[scale=#1]{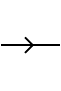}}}
\newcommand{\propagatorD}[1][1.0]{	\raisebox{-.35\height}{\includegraphics[scale=#1]{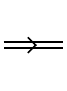}}}
\newcommand{\vertexNNp}[1][1.0]{\raisebox{-.35\height}{\includegraphics[scale=#1]{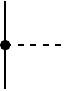}}}
\newcommand{\vertexNNpp}[1][1.0]{\raisebox{-.35\height}{\includegraphics[scale=#1]{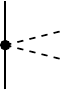}}}
\newcommand{\vertexNDp}[1][1.0]{\raisebox{-.35\height}{\includegraphics[scale=#1]{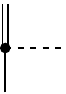}}}
\newcommand{\vertexDNp}[1][1.0]{\raisebox{-.35\height}{\reflectbox{\rotatebox[origin=c]{180}{\includegraphics[scale=#1]{VertexNDp.pdf}}}}}
\newcommand{\vertexDDp}[1][1.0]{\raisebox{-.35\height}{\includegraphics[scale=#1]{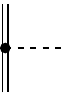}}}
\newcommand{\vertexDDpp}[1][1.0]{\raisebox{-.35\height}{\includegraphics[scale=#1]{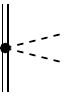}}}

\allowdisplaybreaks

\begin{document}

\begin{frontmatter}

\title{Nucleon-Nucleon Scattering with Coupled Nucleon-Delta Channels in Chiral Effective Field Theory\tnoteref{mytitlenote}}
\tnotetext[mytitlenote]{Work supported in part by DFG and NSFC (CRC110)}

\author[]{Susanne Strohmeier}
\ead{susanne.strohmeier@tum.de}

\author[]{Norbert Kaiser}
\ead{nkaiser@ph.tum.de}

\address{Physik Department, Technische Universität München, 85748 Garching, Germany}

\begin{abstract}
	In this work the elastic scattering of two nucleons is calculated in chiral effective field theory at next-to-leading order taking into account the coupled N$\Delta$-, $\Delta$N- and $\Delta\Delta$-channels. To solve the coupled channel scattering equation one needs as input the potentials for all combinations of these initial and final states. Up to next-to leading order these (transition) potentials arise from one-pion exchange, two-pion exchange and contact interactions. For the two-pion exchange we give analytic expressions for the spectral functions derived from all contributing one-loop diagrams. The forms of the contact potentials at leading and next-to-leading order are determined. 
	We perform a fit of the low energy constants, that belong to the $NN \rightarrow NN$ contact potential and contribute up to next-to-leading order to $S$- and $P$-waves of NN scattering only. The influence of the $\Delta$-isobar dynamics entering through the coupled channels is studied in detail.
\end{abstract}


\end{frontmatter}


\section{Introduction}
\label{sec:introduction} 

In the modern approach to the nuclear force proposed by Weinberg \cite{Weinberg1990Nuclearforcesfrom,Weinberg1991EffectivechiralLagrangians}, the nucleon-nucleon potential is calculated perturbatively within chiral effective field theory and after some regularization this potential is then iterated to all orders by an appropriate scattering equation.

Using dimensional regularization the chiral NN potential has been derived up to next-to-next-to-leading order (N2LO) in Ref.~\cite{Kaiser1997Peripheralnucleonnucleon}, and the $2\pi$-exchange with single and double excitation of virtual $\Delta$-isobars has been considered in Ref.~\cite{Kaiser1998PeripheralNNscattering}.
In Refs.~\cite{Epelbaum1998Nuclearforcesfrom,Epelbaum2000Nuclearforcesfrom} the two nucleon potential was calculated up to N2LO with the method of unitary transformations, including also the $\Delta$-isobar as an explicit degree of freedom in intermediate states for the two-pion exchange.
Further improvements of the chiral NN potential, mostly without the $\Delta$-isobar, were derived by different groups e.g. in Refs.~\cite{Kaiser2001Chiral2pi,Kaiser2002Chiral2pi,Epelbaum2005Twonucleonsystem,Entem2003Accuratechargedependent,Epelbaum2015Improvedchiralnucleon,Epelbaum2015Precisionnucleonnucleon,Entem2015Peripheralnucleonnucleon} going up to the dominant N5LO contributions in Ref.~\cite{Entem2015Dominantcontributionsnucleon}, or considering the N4LO potential including N5LO contact interactions in Ref.~\cite{Reinert2017Semilocalmomentumspace}.

A coupled N$\Delta$-channel approach was used for deriving phenomenological potentials in Refs.~\cite{Bulla1992InclusionpiN,Poepping1987TwoNucleonSystem,Sauer2014ThreeNucleonForces}, and further developed as an extension of the high-precision CD-Bonn potential \cite{Machleidt2001Highprecisioncharge} in Refs.~\cite{Deltuva2003Realistictwobaryon,Deltuva2003ThreeNucleonHadronicElectromagnetic}. 
The construction of the effective chiral Lagrangian for pions, nucleons and $\Delta$-isobars has been initiated in Ref.~\cite{Hemmert1998ChiralLagrangiansand} and continued to higher orders in Ref.~\cite{Fettes2001Pionnucleonscattering} together with specific applications to elastic pion-nucleon scattering. For a recent calculation of NN phase shifts in the SU(3) chiral quark model, treating the coupled octet-decuplet two-baryon channels, see Ref.~\cite{Huang2018Nucleonnucleoninteraction}.

In this work, we include the coupled (N$\Delta$, $\Delta$N, $\Delta\Delta$) channels in the calculation of the S-matrix for elastic nucleon-nucleon scattering in order to investigate the dynamical influence of the strongly coupled $\Delta$-isobar. For this purpose, we derive in chiral effective field theory the interaction potential among nucleons and $\Delta$-isobars at leading and next-to-leading order and make use of a coupled channel scattering equation.

Our paper is organized as follows. 
In \cref{sec:OPE_TPE} we introduce first the prerequisites of our calculation with coupled (NN, N$\Delta$, $\Delta$N, $\Delta\Delta$) channels, give the one-pion exchange potentials and explain in detail the evaluation of the one-loop $2\pi$-exchange diagrams through their spectral functions together with the identification of irreducible parts from planar box diagrams. We give complete expressions for the (transition) potentials as they arise from about 60 one-loop diagrams and determine the structure of the contact potentials including external deltas up to next-to-leading order. Moreover, the method of local regularization for the chiral potentials and the partial wave decomposition for arbitrary spin $s$ are summarized.
In \cref{sec:Kadyshevsky_eq_phase_shifts} we present the Kadyshevsky scattering equation for coupled channels, the nucleon-nucleon K-matrix and the expressions for phase shifts and mixing angles. 
In \cref{sec:results} the results of our calculation with the coupled (NN, N$\Delta$, $\Delta$N, $\Delta\Delta$) channels are presented and discussed, first for the peripheral partial waves with $\ell=3,4,5,6$, and then for the central partial waves. The nine low-energy constants belonging to the nucleon-nucleon contact potential are determined by fitting $S$- and $P$-wave phase shifts to the Nijmegen PWA \cite{Stoks1993Partialwaveanalysis}, considering a cutoff range $\Lambda=400\dots\SI{700}{MeV}$. We also briefly touch upon the deuteron properties.
In \cref{sec:N2LO} we compare the $\Delta$-less theory for the chiral NN potential at N2LO to our coupled (NN, N$\Delta$, $\Delta$N, $\Delta\Delta$) channel approach.
Finally, \cref{sec:conclusion} ends with a summary and conclusions.
In \cref{sec:spin_and_isospin_matrices} we give the explicit form of the spin and isospin matrices for N and $\Delta$, and list relations for products of two such matrices. 
In \cref{sec:coefficients_imaginary_parts} the imaginary parts of the loop functions arising from the $2\pi$-exchange box diagrams are given in analytical form, and
\cref{sec:delta_contact_nlo} contains the next-to-leading order contact potentials with at least one $\Delta$-isobar in the initial or final state.

\section{Nucleon and delta potentials in chiral effective field theory}
\label{sec:OPE_TPE}

Let us start with a general consideration. 
The sixteen possible transition potentials between the four coupled channels (NN, N$\Delta$, $\Delta$N, $\Delta\Delta$) get reduced to ten as a consequence of time reversal symmetry. Among these ten transition potentials the mirror combinations ($N\Delta\rightarrow N \Delta$, $\Delta N\rightarrow  \Delta N$), ($NN\rightarrow N \Delta$, $NN\rightarrow \Delta N$) and ($N\Delta\rightarrow \Delta \Delta$, $\Delta N\rightarrow \Delta\Delta$) lead to identical matrix elements, so that one arrives at seven independent transitions.

The calculation of these transition potentials in chiral effective field theory up to NLO consists of evaluating one-pion exchange tree-diagrams and two-pion exchange loop-diagrams with the propagators and vertices given in \cref{tb:propagators_vertices}. They stem from the chiral Lagrangians in the heavy baryon limit, with $v^{\mu} = (1, \vec 0)$ the four-velocity and $\Delta=\SI{293}{MeV}$ the delta-nucleon mass splitting. Furthermore, $\tau^{a}$ and $\vec \sigma$ are the usual (isospin and spin) $2\times 2$ Pauli matrices, $T^{a}$ and $\vec S$ are respective transition operators and $\Theta^{a}$ and $\vec \Sigma$ are $4\times 4$ isospin and spin  matrices for the delta. Their explicit form together with various relations for products are given in \cref{sec:spin_and_isospin_matrices}. The vertices in \cref{tb:propagators_vertices} are written with momenta of the in- and outgoing pions denoted by $q$.
\begin{table}[!htb]
	\centering
	\caption{Propagators of nucleons and deltas and leading order chiral vertices for pionic couplings in the heavy baryon approach.}
	\label{tb:propagators_vertices}
	\begin{tabular}{ccc}
		\toprule
		$N$ & \propagatorN[0.75] & $\dfrac{\iu}{v \cdot l + \iu \epsilon}$\\ [10pt]
		$\Delta$ & \propagatorD[0.75] & $\dfrac{\iu}{v \cdot l - \Delta + \iu \epsilon}$\\ [10pt]
		\midrule 
		$N\rightarrow\pi ^{a} N$ & \vertexNNp[0.75] & $-\dfrac{\ga}{2 \fpi} \vec \sigma \cdot \vec q \,\tau^{a}$\\ [10pt]
		$N  \rightarrow\pi ^{a} \Delta$ & \vertexNDp[0.75] & $-\dfrac{3\ga}{2\sqrt{2} \fpi} \vec S^{\dagger} \cdot \vec q \, T^{a\dagger}$\\ [10pt]
		$\Delta  \rightarrow\pi ^{a} N$ & \vertexDNp[0.75] & $-\dfrac{3\ga}{2\sqrt{2} \fpi} \vec S \cdot \vec q \, T^{a}$\\ [10pt]
		$\Delta \rightarrow\pi ^{a} \Delta$ & \vertexDDp[0.75] & $-\dfrac{\ga}{10 \fpi} \vec \Sigma \cdot \vec q \, \Theta^{a}$\\ [10pt]
		$\pi^{b} N  \rightarrow\pi ^{a} N$ & \vertexNNpp[0.75] & $\dfrac{1}{4 \fpi^2} \epsilon^{bac} \tau^{c}  v \cdot ( q _{\pi^a} + q_{\pi^b})$\\ [10pt]
		$\pi^{b} \Delta  \rightarrow\pi ^{a} \Delta$ & \vertexDDpp[0.75] & $\dfrac{1}{4 \fpi^2} \epsilon^{bac} \Theta^{c}  v \cdot ( q _{\pi^a} + q_{\pi^b})$\\ [10pt]
		\bottomrule
	\end{tabular}
\end{table}
For the $\pi $N$\Delta$ and $\pi \Delta \Delta$ coupling constants the large $N_C$-relations $g_{\pi N  \Delta} = 3 g_{\pi NN}/\sqrt{2}$ and $g_{\pi \Delta \Delta}= g_{\pi NN}/5$ have been used together with the Goldberger-Treiman relation $g_{\pi NN}=\ga \mnuc /\fpi$, where $\fpi=\SI{92.4}{MeV}$ denotes the weak pion decay constant. The nucleon axial vector coupling is chosen here as $\ga=1.29$ in order to obtain the value $g_{\pi NN}^2/(4\pi)=13.6$, that is consistent with a recent determination from $\pi N$-scattering data based on the Goldberger-Miyazawa-Oehme sum rule $g_{\pi N}^2/(4\pi)=13.69\pm 0.20$ \cite{Baru2011Precisioncalculation}.
The last two vertices in \cref{tb:propagators_vertices} are the Tomozawa-Weinberg $2\pi$-contact vertices for nucleons and deltas, fixed by chiral symmetry. At this point one should note that its off-diagonal (N$\Delta$) counterpart is of higher order and thus not relevant for our NLO calculation.

\begin{figure}[!htb]
	\centering
	\includegraphics[width=\textwidth]{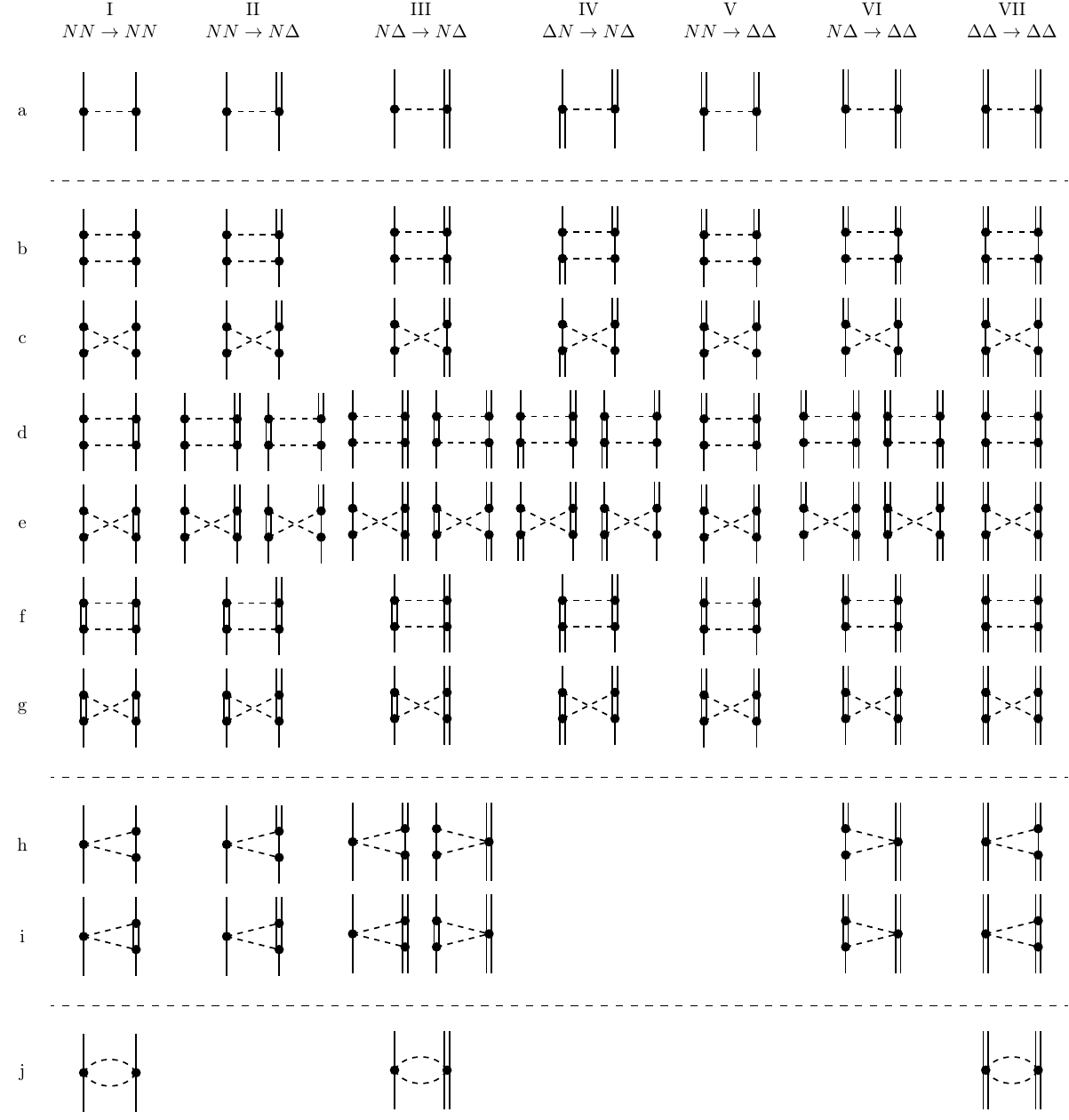}
	\caption{One-pion and two-pion exchange diagrams contributing to the (transition) potentials for the coupled NN-, N$\Delta$-, $\Delta$N- and $\Delta\Delta$-channels up to next-to-leading order.}
	\label{fig:OPE_TPE_diagrams}
\end{figure}

\subsection{One-pion exchange}
\label{subsec:OPE} 
\noindent
The one-pion exchange potentials arising from the tree diagrams in line (a) of \cref{fig:OPE_TPE_diagrams} take the following form,
\begin{align}
V_{NN \rightarrow NN}^{OPE} &= \frac{\ga^2 (\vec \sigma_1 \cdot \vec q \,) (\vec \sigma_2 \cdot \vec q \,)}{4 \fpi^2 (q^2 + \mpi^2)} \vec \tau_1 \cdot \vec \tau_2 \label{eq:OPE_NNNN}\\
V_{NN \rightarrow N\Delta}^{OPE} &= \frac{3\ga^2 (\vec \sigma_1 \cdot \vec q \,) (\vec S_2^{\dagger} \cdot \vec q \,)}{4\sqrt{2} \fpi^2 (q^2 + \mpi^2)} \vec \tau_1 \cdot \vec T_2^{\dagger}  \label{eq:OPE_NNND}\\
V_{N\Delta \rightarrow N\Delta}^{OPE} &= \frac{\ga^2 (\vec \sigma_1 \cdot \vec q \,) (\vec \Sigma_2 \cdot \vec q \,)}{20 \fpi^2 (q^2 + \mpi^2)} \vec \tau_1 \cdot \vec \Theta_2  \label{eq:OPE_NDND}\\
V_{\Delta N \rightarrow N\Delta}^{OPE} &= \frac{9\ga^2 (\vec S_1 \cdot \vec q \,) (\vec S_2^{\dagger} \cdot \vec q \,)}{8 \fpi^2 (q^2 + \mpi^2)} \vec T_1 \cdot \vec T_2^{\dagger}  \label{eq:OPE_DNND}\\
V_{NN \rightarrow \Delta\Delta}^{OPE} &= \frac{9\ga^2 (\vec S_1^{\dagger} \cdot \vec q \,) (\vec S_2^{\dagger} \cdot \vec q \,)}{8 \fpi^2 (q^2 + \mpi^2)} \vec T_1^{\dagger} \cdot \vec T_2^{\dagger}  \label{eq:OPE_NNDD}\\
V_{N\Delta \rightarrow \Delta\Delta}^{OPE} &= \frac{3\ga^2 (\vec S_1^{\dagger} \cdot \vec q \,) (\vec \Sigma_2 \cdot \vec q \,)}{20\sqrt{2} \fpi^2 (q^2 + \mpi^2)} \vec T_1^{\dagger} \cdot \vec \Theta_2  \label{eq:OPE_NDDD}\\
V_{\Delta\Delta \rightarrow \Delta\Delta}^{OPE} &= \frac{\ga^2 (\vec \Sigma_1 \cdot \vec q \,) (\vec \Sigma_2 \cdot \vec q \,)}{100 \fpi^2 (q^2 + \mpi^2)} \vec \Theta_1 \cdot \vec \Theta_2 \label{eq:OPE_DDDD}
\end{align} 
where $\mpi=\SI{138.03}{MeV}$ denotes the average pion mass, and the indices 1 and 2 refer to the left and right baryon line, respectively.
In the (only) relevant channels with total isospin $I=0,1$ the isospin operators at the end of \cref{eq:OPE_NNNN,eq:OPE_NNND,eq:OPE_NDND,eq:OPE_DNND,eq:OPE_NNDD,eq:OPE_NDDD,eq:OPE_DDDD} have the eigenvalues as given in \cref{tb:Isospin_OPE}.
\begin{table}[!htb]
	\caption{Isospin factors of the one-pion exchange potentials for total isospin $I=0,1$}
	\label{tb:Isospin_OPE}
	\centering
	\begin{tabular}{*{7}{c}}
		\toprule
		  $ NN \! \rightarrow \! NN$ &
		  $ NN \! \rightarrow \! N\Delta$ &
		  $ N\Delta \! \rightarrow \! N\Delta$ &
		  $ \Delta N \! \rightarrow \! N\Delta$ &
		  $ NN \! \rightarrow \! \Delta\Delta$ &
		  $ N\Delta \! \rightarrow \! \Delta\Delta$ &
		  $ \Delta\Delta \! \rightarrow \! \Delta\Delta$  \\
 		  $4I-3$ & $2\sqrt{\frac{2}{3}}I$ & $-5I$ & $-\frac{1}{3}I$ & $-\frac{3(1\! - \! I) + \sqrt{5} I}{3\sqrt{2}}$ & $-2\sqrt{\frac{5}{3}}I$ & $4I-15$ \\ 
		\bottomrule
	\end{tabular} 
\end{table}

\subsection{Two-pion exchange}
\label{subsec:TPE} 
\noindent
\begin{figure}[!htb]
	\centering
	\includegraphics[scale=1.0]{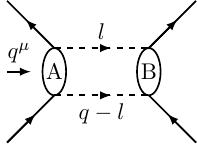}
	\caption{Schematic figure of a two-pion exchange process between two baryons with momentum transfer $q$ and loop momentum $l$.}
	\label{fg:calculation_of_potentials}
\end{figure}
For the determination of the two-pion exchange potentials we follow the approach of Ref.~\cite{Kaiser2001Chiral2pi} by first calculating the imaginary parts of one-loop diagrams with the help of the Cutkosky cutting rules \cite{Cutkosky1960SingularitiesdiscontinuitiesFeynman}. For a generic two-pion exchange process between two baryons as shown in \cref{fg:calculation_of_potentials} the corresponding imaginary part is given by the expression
\begin{align}
	\Im \int \frac{\d^4 l}{(2\pi)^4 \iu} \frac{\iu^2}{(l^2-\mpi^2)((q-l)^2-\mpi^2)} A \cdot B = -\frac{1}{2} \int \d \Phi A\cdot B \; ,
	\label{eq:imaginary_part}
\end{align}
where $A$ and $B$ denote S-matrices for the left and right $2\pi$-baryon interaction.
These imaginary parts refer to an analytical continuation of the loop-amplitude to timelike momentum transfers $q \! \cdot \! q = \mu^2 > 4 \mpi^2$, corresponding formally to $ \abs{ \vec q \,}=\iu \mu + 0^{+}$. The right hand side of \cref{eq:imaginary_part} denotes an integral of the baryon-antibaryon $\rightarrow 2\pi \rightarrow$ baryon-antibaryon transition amplitude over the Lorentz-invariant $2\pi$-phase space, which is most conveniently worked out in the $2\pi$ center-of-mass frame, where it reduces to an angular integral of the form 
\begin{align}
	\int \d \Phi \, F  = \frac{\sqrt{\mu^2-\mpi^2}}{16 \pi \mu} \int_{-1}^{1}\limits \d x \, F  \; .
\end{align}
Here, $F$ represents a generic function depending on $\mu$ and $x$.

The phase space integrals arising from the two-pion exchange diagrams collected in \cref{fig:OPE_TPE_diagrams} can be split into terms that contain up to four powers of the loop momentum $\vec l$, which are contracted with spin operators. For each such term we extend $\vec l$ to $l^{\mu}$ and determine the corresponding Lorentz tensor structure. 
The first case concerns one power of the loop momentum $l^{\mu}$. The corresponding $2\pi$-phase space integral has the form 
\begin{align}
\int \d \Phi \, F \,  l^{\mu} =& q^{\mu} \widetilde A_{1} + v^{\mu}  \widetilde B_{1} \; ,
\label{eq:phase_space_integral_l_1}
\end{align}
where $F$ stands for further factors coming from baryon propagators and pionic vertices. The second Lorentz vector $v^{\mu}$ on the right hand side produces a vanishing result when contracted with spin-operators. 
By applying the projector $ q_{\mu}/\mu^2 $ on \cref{eq:phase_space_integral_l_1} and using the relations $v\cdot q=0$ and $l \cdot q = \mu^2/2$ one obtains the following formula to calculate the coefficient of interest
\begin{align}
\widetilde A_{1}=\int \d \Phi \, \frac{F}{2} \; .
\label{eq:coefficients_1_l}
\end{align}
For two powers of the loop momentum $l^{\mu}$ we decompose the second-rank Lorentz tensor resulting from the phase space integration as
\begin{align}
\int \d \Phi \, F  \, l^{\mu}l^{\nu} =& -g^{\mu  \nu } \widetilde A_{2} + q^{\mu } q^{\nu } \widetilde B_{2} + \dots \; ,
\end{align}
where the dots on the right hand side stand for two tensor terms with $v^{\mu}$ or  $v^{\nu}$, that are not needed.
After solving four linear equations, one finds for the two coefficients of interest the following expressions
\begin{align}
\widetilde A_{2}=&\int \d \Phi \,  \frac{F}{8} \left(x^2-1\right) \left(4 \mpi^2-\mu ^2\right) \; , \label{eq:coefficients_2_l_a} \\
\widetilde B_{2}=& \int \d \Phi \, F   \frac{4 \mpi^2 \left(x^2-1\right)-\mu ^2 \left(x^2-3\right)}{8 \mu ^2} \; .
\label{eq:coefficients_2_l_b}
\end{align}
The phase space integral including three powers of the loop momentum is decomposed as
\begin{align}
\int \d \Phi \, F \,  l^{\mu}l^{\nu}l^{\rho} =&  \left(-q^{\rho } g^{\mu  \nu }-q^{\nu } g^{\mu  \rho }-q^{\mu } g^{\nu  \rho }\right) \widetilde A_{3} + q^{\mu } q^{\nu } q^{\rho }\widetilde B_{3} + \dots \; ,
\end{align}
where the dots on the right-hand side stand for four tensor terms with $v^{\mu}$, $v^{\nu}$ or $v^{\rho}$.
The two coefficients of interest are calculated as follows
\begin{align}
\widetilde A_{3}=& \int \d \Phi \, \frac{F}{16} \left(x^2-1\right) \left(4 \mpi^2-\mu ^2\right) \; ,\\
\widetilde B_{3}=& \int \d \Phi \, F \frac{12 \mpi^2 \left(x^2-1\right)+\mu ^2 \left(5-3 x^2\right)}{16 \mu ^2} \; .
\label{eq:coefficients_3_l}
\end{align}
The phase space integral with four loop momenta has the following Lorentz structure, 
\begin{align}
\int \d \Phi \, F  \, l^{\mu}l^{\nu}l^{\rho}l^{\sigma} =& \left(g^{\mu  \nu } g^{\rho  \sigma }+g^{\mu  \rho } g^{\nu  \sigma }+g^{\mu  \sigma } g^{\nu  \rho }\right) \widetilde A_{4} \nonumber \\
+& \left(-q^{\rho } q^{\sigma } g^{\mu  \nu } \!\!-\! q^{\nu } q^{\sigma } g^{\mu  \rho } \!\!-\! q^{\mu } q^{\sigma } g^{\nu  \rho } \!\!-\! q^{\nu } q^{\rho } g^{\mu  \sigma } \!\!-\! q^{\mu } q^{\rho } g^{\nu  \sigma } \!\!-\! q^{\mu } q^{\nu } g^{\rho  \sigma }\right) \widetilde B_{4} \nonumber\\
+& \; q^{\mu } q^{\nu } q^{\rho } q^{\sigma } \widetilde C_{4} + \dots \; .
\end{align}
Out of the nine coefficients we find for the three coefficients of interest the following expressions
\begin{align}
\widetilde A_{4}=& \int \d \Phi \,  \frac{F}{128} \left(x^2-1\right)^2 \left(\mu ^2-4 \mpi^2\right)^2 \; ,\\
\widetilde B_{4}=& \int \d \Phi \, F \frac{\left(x^2-1\right) \left(\mu ^2-4 \mpi^2\right) \left[ \mu ^2 \left(x^2-5\right)-4 \mpi^2 \left(x^2-1\right)\right]}{128 \mu ^2} \; , \\
\widetilde C_{4}=& \int \d \Phi \, F  \frac{48 \mpi^4 \left(x^2-1\right)^2-24 \mu ^2 \mpi^2 \left(x^4-6 x^2+5\right)+\mu ^4 \left(3 x^4-30 x^2+35\right)}{128 \mu ^4} \; .
\label{eq:coefficients_4_l}
\end{align}

\subsubsection{Box diagrams}
\label{sec:box}
The set of planar and crossed box diagrams is shown in lines (b) to (g) of \cref{fig:OPE_TPE_diagrams}.
Reducible parts of planar box diagrams will be generated by iteration of the $1\pi$-exchange potentials in the coupled channel scattering equation, and therefore they have to be excluded from the $2\pi$-exchange potentials. In order to identify the irreducible parts of the planar box diagrams we calculate the $l_0$-integral over the pertinent products of heavy baryon and pion propagators. Denoting the on-shell energies of the two exchanged pions by $\omega_{1}$ and $\omega_{2}$, and applying residue calculus, we find the following results for planar and crossed box diagrams with intermediate N$\Delta$ and $\Delta\Delta$ states:

\begin{align}
\diagramboxDmirrored ~\longrightarrow &~ \int \frac{\d l_0}{2 \pi \iu} \frac{1}{(l_0 -\Delta +\iu \epsilon )(-l_0 +\iu \epsilon) (l_0^2 -\omega_{1}^2 + \iu \epsilon)(l_0^2 -\omega_{2}^2 + \iu \epsilon)} \nonumber\\
&= 
\frac{1}{\Delta \omega_{1}^2 \omega_{2}^2} - \frac{\omega_{1}^{2} + \omega_{1} \omega_{2} + \omega_{2}^{2}+\Delta( \omega_{1} +\omega_{2})}{2 \omega_{1}^{2}\omega_{2}^{2}( \omega_{1} +\omega_{2})(\omega_{1}+ \Delta) (\omega_{2} +\Delta)} \; ,
\\
\diagramboxcrossedDmirrored ~\longrightarrow &~ \int \frac{\d l_0}{2 \pi \iu} \frac{1}{(l_0 -\Delta +\iu \epsilon )(l_0 +\iu \epsilon) (l_0^2 -\omega_{1}^2 + \iu \epsilon)(l_0^2 -\omega_{2}^2 + \iu \epsilon)}  \nonumber\\
&=  
\frac{\omega_{1}^{2} + \omega_{1} \omega_{2} + \omega_{2}^{2}+\Delta( \omega_{1} +\omega_{2})}{2 \omega_{1}^{2}\omega_{2}^{2}( \omega_{1} +\omega_{2})(\omega_{1}+ \Delta) (\omega_{2} +\Delta)} \; , \label{eq:irreducible_ND_crossed}
\\
\diagramboxDD ~\longrightarrow &~ \int \frac{\d l_0}{2 \pi \iu} \frac{1}{(l_0 -\Delta +\iu \epsilon )(-l_0 -\Delta +\iu \epsilon) (l_0^2 -\omega_{1}^2 + \iu \epsilon)(l_0^2 -\omega_{2}^2 + \iu \epsilon)} \nonumber\\
&= 
\frac{1}{2\Delta \omega_{1}^2 \omega_{2}^2} - \frac{\omega_{1}^{2} + \omega_{1} \omega_{2} + \omega_{2}^{2}+\Delta( \omega_{1} +\omega_{2})}{2 \omega_{1}^{2}\omega_{2}^{2}( \omega_{1} +\omega_{2})(\omega_{1}+ \Delta) (\omega_{2} +\Delta)} \; , \\
\diagramboxcrossedDD ~\longrightarrow &~ \int \frac{\d l_0}{2 \pi \iu} \frac{1}{(l_0 -\Delta +\iu \epsilon )^2 (l_0^2 -\omega_{1}^2 + \iu \epsilon)(l_0^2 -\omega_{2}^2 + \iu \epsilon)} \nonumber\\
&=  
\frac{\omega_{1}^{2} + \omega_{1} \omega_{2} + \omega_{2}^{2}+2\Delta( \omega_{1} +\omega_{2})+\Delta^2}{2 \omega_{1}\omega_{2}( \omega_{1} +\omega_{2})(\omega_{1}+ \Delta)^2 (\omega_{2} +\Delta)^2} \; . \label{eq:irreducible_DD_crossed}
\end{align}
The reducible parts are identified by the $1/ \Delta$-dependence, since $\Delta$ or $2\Delta$ is the remaining energy denominator when neglecting kinetic energies. 
By comparison with \cref{eq:irreducible_ND_crossed}, one makes the interesting observation that the irreducible parts of the planar N$\Delta$ and $\Delta\Delta$ boxes are \textit{equal}, and coincide with the \textit{negative} of the crossed N$\Delta$ box. Note that the crossed $\Delta\Delta$ box in \cref{eq:irreducible_DD_crossed} has a completely different structure.
At that point we remind that the irreducible part of the planar NN box is in the same way equal to the \textit{negative} of the crossed NN box, as shown in Ref.~\cite{Kaiser1997Peripheralnucleonnucleon}.

In the presence of external $\Delta$-isobars the spin structure of the two-pion exchange potential gets more complicated. A simple reduction of spin operators through the relations $\sigma^{i}\sigma^{j}= \delta^{ij} \1+ \iu \epsilon^{ijk} \sigma^{k}$ and $S^{i} S^{j \dagger} = \frac{1}{3}\bigl(2 \delta^{ij} \1- \iu \epsilon^{ijk} \sigma^{k} \bigr)$, which is convenient for the NN potential, does not exist for the $\Delta$-sector.
To treat this problem in an efficient way, we split for the box diagrams the product of spin matrices $\mathcal{S}_{2}^{lk}$ and the product of isospin matrices $\mathcal{T}_{2}^{ji}$ belonging to the second baryon into a symmetric and an antisymmetric part (under the exchange of the upper indices).
In combination with the spin and isospin operators of the first baryon one can construct the even and odd operators
\begin{align}
\mathcal{S}^{+}&=\mathcal{S}_{1}^{ij} \left(\frac{1}{2}\mathcal{S}_{2}^{kl}+\frac{1}{2}\mathcal{S}_{2}^{lk} \right) \;, \label{eq:decomposition_spin_isospin_1}\\*
\mathcal{S}^{-}&=\mathcal{S}_{1}^{ij} \left(\frac{1}{2}\mathcal{S}_{2}^{kl}-\frac{1}{2}\mathcal{S}_{2}^{lk} \right) \;, \label{eq:decomposition_spin_isospin_2}\\*
\mathcal{T}^{+}&=\mathcal{T}_{1}^{ij} \left(\frac{1}{2}\mathcal{T}_{2}^{ij}+\frac{1}{2}\mathcal{T}_{2}^{ji} \right) \;, \label{eq:decomposition_spin_isospin_3}\\*
\mathcal{T}^{-}&=\mathcal{T}_{1}^{ij} \left(\frac{1}{2}\mathcal{T}_{2}^{ij}-\frac{1}{2}\mathcal{T}_{2}^{ji} \right) \;.\label{eq:decomposition_spin_isospin_4}
\end{align}
By making use of these definitions and merging planar and crossed box diagrams, the imaginary part of the two-pion exchange potential $V$ from box diagrams can be split into four different parts
\begin{align}
\Im V^{++}&=\mathcal{T}^{+}\mathcal{S}^{+} \Im \left(\diagrambox +\diagramboxcrossed \right), \label{eq:decomposition_potential_1}\\*
\Im V^{+-}&=\mathcal{T}^{+}\mathcal{S}^{-} \Im \left(\diagrambox -\diagramboxcrossed \right), \label{eq:decomposition_potential_2}\\*
\Im V^{-+}&=\mathcal{T}^{-}\mathcal{S}^{+} \Im \left(\diagrambox -\diagramboxcrossed \right), \label{eq:decomposition_potential_3}\\*
\Im V^{--}&=\mathcal{T}^{-}\mathcal{S}^{-} \Im \left(\diagrambox +\diagramboxcrossed \right), \label{eq:decomposition_potential_4}
\end{align}
where the symbolic diagrams in parenthesis represent the irreducible part of the planar box and the crossed box without their spin and isospin operators. Note that the sign combination on $V^{\pm\pm}$ tells whether these diagrams are added or subtracted.

\begin{table}[!htb]
	\centering
	\caption{Isospin factors of box diagrams evaluated for the total isospin $I=0,1$ as described in the text.}
	\label{tb:Isospinfactors}
	\begin{tabular}{r*{5}{c}}
		\toprule
		& $ NN \rightarrow NN$  & $ NN \rightarrow N\Delta$ & $ N\Delta \rightarrow N\Delta$ & $ \Delta N \rightarrow N\Delta$  \\ \midrule
		$\mathcal{T}^{+ NN}$			& $3$    & $0$ & $1$ & $-\frac{5}{2}I$  \\[5pt]
		$\mathcal{T}^{- NN}$			& $6-8I$ & $2\sqrt{\frac{2}{3}}I$ & $\frac{5}{3}I$ & $\frac{1}{6}I$  \\[5pt]
		$\mathcal{T}^{+ N\Delta}$		& $2$    & $0$ & $15$ & $-\frac{5}{2}I$  \\[5pt]
		$\mathcal{T}^{- N\Delta}$		& $-2+\frac{8}{3}I$  & $-10\sqrt{\frac{2}{3}}I$ & $10I$ & $-\frac{5}{6}I$  \\[5pt]
		$\mathcal{T}^{+ \Delta N}$		& $2$ & $0$ & $\frac{2}{3}$ & $-\frac{5}{2}I$ \\[5pt]
		$\mathcal{T}^{- \Delta N}$		& $-2+\frac{8}{3}I$ & $-\frac{2}{3}\sqrt{\frac{2}{3}}I$ & $-\frac{5}{9}I$ & $-\frac{5}{6}I$  \\[5pt]
		$\mathcal{T}^{+ \Delta\Delta}$	& $\frac{4}{3}$ & $0$ & $10$ & $-\frac{5}{2}I$  \\[5pt]
		$\mathcal{T}^{- \Delta\Delta}$	& $\frac{2}{3}-\frac{8}{9}I$ & $\frac{10}{3}\sqrt{\frac{2}{3}}I$ & $-\frac{10}{3}I$ & $\frac{25}{6}I$  \\[5pt] \bottomrule
		\toprule
		& \multicolumn{2}{c}{$ NN \rightarrow \Delta\Delta$} & $ N\Delta \rightarrow \Delta\Delta$ & $ \Delta\Delta \rightarrow \Delta\Delta$ \\ \midrule
		$\mathcal{T}^{+ NN}$			&  \multicolumn{2}{c}{$\frac{1}{\sqrt{2}}(5(1\! - \! I)-\sqrt{5}I)$} & $-\sqrt{\frac{5}{3}}I$ & $\frac{1}{6}(7-4I)$ \\[5pt]
		$\mathcal{T}^{- NN}$			&  \multicolumn{2}{c}{$\frac{1}{3\sqrt{2}}( 3(1\! - \! I) + \sqrt{5} I)$} & $-\frac{1}{3} \sqrt{\frac{5}{3}} I$ & $\frac{5}{6} - \frac{2}{9} I$ \\[5pt]
		$\mathcal{T}^{+ N\Delta}$		&  \multicolumn{2}{c}{$\frac{1}{\sqrt{2}}(5(1\! - \! I) - \sqrt{5}I)$} & $2 \sqrt{10} I$ & $-5+8I$ \\[5pt]
		$\mathcal{T}^{- N\Delta}$		&  \multicolumn{2}{c}{$-\frac{5}{3\sqrt{2}}(3(1\! - \! I) + \sqrt{5} I)$} & $-2\sqrt{\frac{5}{3}} I$ & $5-\frac{4}{3}I$ \\[5pt]
		$\mathcal{T}^{+ \Delta N}$		&  \multicolumn{2}{c}{$\frac{1}{\sqrt{2}}(5(1\! - \! I)-\sqrt{5}I)$} & $-\sqrt{\frac{5}{3}} I$ & $-5+8I$ \\[5pt]
		$\mathcal{T}^{- \Delta N}$		&  \multicolumn{2}{c}{$-\frac{5}{3\sqrt{2}}( 3(1\! - \! I)+\sqrt{5} I)$} & $\frac{5}{3}\sqrt{\frac{5}{3}}I$ & $5-\frac{4}{3}I$ \\[5pt]
		$\mathcal{T}^{+ \Delta\Delta}$	&  \multicolumn{2}{c}{$\frac{1}{\sqrt{2}}(5(1\! - \! I)-\sqrt{5} I)$} & $2\sqrt{10}I$ & $195-96I$ \\[5pt]
		$\mathcal{T}^{- \Delta\Delta}$	&  \multicolumn{2}{c}{$-\frac{25}{3\sqrt{2}}( 3(1\! - \! I)+\sqrt{5} I )$} & $10\sqrt{\frac{5}{3}}I$ & $30-8I$ \\[5pt] \bottomrule
	\end{tabular}
\end{table}
In the following we list the analytical results for the two-pion exchange potentials from box diagrams, ordered by their initial and final states.
In view of their large number, these potentials are labeled as $V_{in~out}^{\pm \pm~int}$, where the two signs~$\pm$ refer to the decompositions in \cref{eq:decomposition_potential_1,eq:decomposition_potential_2,eq:decomposition_potential_3,eq:decomposition_potential_4}. The subscripts $in$ and $out$ denote the two ingoing and outgoing baryons, respectively, and $int$ refers to the intermediate baryon pair. The $q$-dependent functions $A_i$, $B_i$ and $C_i$ have to be calculated numerically as (regularized) dispersion integrals from their imaginary parts, which are collected in \cref{sec:coefficients_imaginary_parts}. The isospin factors $\mathcal{T}_{in~out}^{\pm int}$ in front of the potentials can be calculated with little effort for total isospin $I=0,1$ and are collected in \cref{tb:Isospinfactors}. It is worth mentioning, that (nonvanishing) potentials $V^{++}$ and $V^{--}$ exist only for the $\Delta\Delta$ intermediate state. This is a consequence of the negative sign of the irreducible part from planar NN and N$\Delta$ boxes.

\noindent
One should note that in our one-loop calculation of $2\pi$-exchange potentials external deltas have to be treated kinematically as off-shell particles\footnote{The threshold for real delta production $T_{\text{lab}}=\SI{632}{MeV}$ lies far in the inelastic region.} with energies close to the nucleon mass.

\noindent
\begin{align}
\shortintertext{\underline{I) NN$\rightarrow$NN:}}
V_{NNNN}^{+ - NN}=&   \mathcal{T}_{NNNN}^{+ NN} 
 \frac{\ga^4}{16\fpi^4}  
A_2^{NN-} (\vec q \! \cdot \! \vec{\sigma }_1 \; \vec q \! \cdot \! \vec{\sigma }_2-q^2 \vec{ \sigma }_1 \! \cdot \! \vec{ \sigma }_2)	\nonumber\\ 
V_{NNNN}^{- + NN}=&   \mathcal{T}_{NNNN}^{- NN} 
\frac{\ga^4}{16\fpi^4}   
\biggl\lbrace q^2 \Bigl[A_2^{NN-}\!\! + 10 (B_4^{NN-}\!\! - A_3^{NN-}) \nonumber\\*&
+ q^2 (B_2^{NN-}\!\! - 2 B_3^{NN-}\!\! + C_4^{NN-})\Bigr] + 15 A_4^{NN-} \biggr\rbrace	\nonumber\\ 
V_{NNNN}^{+ - N\Delta}=&   \mathcal{T}_{NNNN}^{+ N\Delta} 
\frac{3\ga^4}{32\fpi^4}   
 A_2^{N\Delta-} (q^2 \vec{ \sigma }_1 \! \cdot \! \vec{ \sigma }_2 -\vec q \! \cdot \! \vec{\sigma }_1 \; \vec q \! \cdot \! \vec{\sigma }_2)	\nonumber\\  
V_{NNNN}^{- +N\Delta}=&   \mathcal{T}_{NNNN}^{- N\Delta} 
\frac{9\ga^4}{32\fpi^4} 
\biggl\lbrace\frac{2}{3} q^2 \Bigl[A_2^{N\Delta-}\!\! + 10 (B_4^{N\Delta-}\!\! - A_3^{N\Delta-}) \nonumber\\*
&+q^2 (B_2^{N\Delta-}\!\! - 2 B_3^{N\Delta-}\!\! + C_4^{N\Delta-})\Bigr] + 10 A_4^{N\Delta-} \biggr\rbrace\nonumber\\ 
V_{NNNN}^{\pm -\Delta\Delta}=&   \mathcal{T}_{NNNN}^{\pm \Delta\Delta} 
\frac{9\ga^4}{64\fpi^4} 
 A_2^{\Delta\Delta\mp} (\vec q \! \cdot \! \vec{\sigma }_1 \; \vec q \! \cdot \! \vec{\sigma }_2-q^2 \vec{ \sigma }_1 \! \cdot \! \vec{ \sigma }_2) \nonumber\\ 
V_{NNNN}^{\pm +\Delta\Delta}=&   \mathcal{T}_{NNNN}^{\pm \Delta\Delta} 
\frac{9\ga^4}{16\fpi^4} 
 \biggl\lbrace q^2 \Bigl[A_2^{\Delta\Delta\pm}\!\! + 10 (B_4^{\Delta\Delta\pm}\!\! - A_3^{\Delta\Delta\pm}) \nonumber\\*
& +q^2 (B_2^{\Delta\Delta \pm}\!\! - 2 B_3^{\Delta\Delta\pm}\!\! + C_4^{\Delta\Delta\pm})\Bigr] + 15 A_4^{\Delta\Delta\pm} \biggr\rbrace &
\label{eq:potentials_NNNN}
\\\nonumber\\
\shortintertext{\underline{II) NN$\rightarrow$N$\Delta$:}}
V_{NNN\Delta}^{+ - NN}=&   \mathcal{T}_{NNN\Delta}^{+ NN}
\frac{3\ga^4}{32\sqrt{2}\fpi^4}  
 A_2^{NN-} (q^2 \vec{ \sigma }_1 \! \cdot \! \vec{ S}_2^{\dagger} - \vec q \! \cdot \! \vec{\sigma }_1 \; \vec q \! \cdot \! \vec{S}_2^{\dagger})	\nonumber\\ 
V_{NNN\Delta}^{- +NN}=&   \mathcal{T}_{NNN\Delta}^{- NN}
\frac{-3\sqrt{3}\ga^4}{32\fpi^4} 
(q^i q^j S_2^{i j \dagger}) \biggl[A_2^{NN -}\!\! - 7 A_3^{NN -}\!\! + 7 B_4^{NN -} \nonumber\\*
&+q^2 (B_2^{NN -}\!\! - 2 B_3^{NN -}\!\! + C_4^{NN -})\biggr]
\nonumber\\ 
V_{NNN\Delta}^{+ -N\Delta }=&   \mathcal{T}_{NNN\Delta}^{+ N\Delta}
\frac{3\ga^4}{32\sqrt{2}\fpi^4} 
A_2^{N\Delta -} (\vec q \! \cdot \! \vec{\sigma }_1 \; \vec q \! \cdot \! \vec{S}_2^{\dagger}-q^2 \vec{ \sigma }_1 \! \cdot \! \vec{ S}_2^{\dagger})	\nonumber\\ 
V_{NNN\Delta}^{- +N\Delta }=&   \mathcal{T}_{NNN\Delta}^{- N\Delta}
\frac{-\sqrt{3}\ga^4}{80\fpi^4}  
 (q^i q^j S_2^{i j \dagger}) \biggl[A_2^{N\Delta -}\!\! - 7 A_3^{N\Delta -}\!\! + 7 B_4^{N\Delta -} \nonumber\\*
&+q^2 (B_2^{N\Delta -}\!\! - 2 B_3^{N\Delta -}\!\! +C_4^{N\Delta -})\biggr]
\nonumber\\ 
V_{NNN\Delta}^{+ -\Delta N }=&   \mathcal{T}_{NNN\Delta}^{+ \Delta N}
\frac{9\ga^4}{64\sqrt{2}\fpi^4}   
 A_2^{N\Delta -} (\vec q \! \cdot \! \vec{\sigma }_1 \; \vec q \! \cdot \! \vec{S}_2^{\dagger}-q^2 \vec{ \sigma }_1 \! \cdot \! \vec{ S}_2^{\dagger})	\nonumber\\
V_{NNN\Delta}^{- +\Delta N }=&   \mathcal{T}_{NNN\Delta}^{- \Delta N}
\frac{-27\sqrt{3}\ga^4}{64\fpi^4}  
 (q^i q^j S_2^{i j \dagger}) \biggl[A_2^{N\Delta -}\!\! - 7 A_3^{N\Delta -}\!\! + 7 B_4^{N\Delta -} \nonumber\\*
&+q^2 (B_2^{N\Delta -}\!\! - 2 B_3^{N\Delta -}\!\! + C_4^{N\Delta -})\biggr]\nonumber\\
V_{NNN\Delta}^{\pm -\Delta\Delta }=&   \mathcal{T}_{NNN\Delta}^{\pm \Delta\Delta}
\frac{9\ga^4}{64\sqrt{2}\fpi^4}   
A_2^{\Delta\Delta \mp} (q^2 \vec{ \sigma }_1 \! \cdot \! \vec{ S}_2^{\dagger} -\vec q \! \cdot \! \vec{\sigma }_1 \; \vec q \! \cdot \! \vec{S}_2^{\dagger})	\nonumber\\ 
V_{NNN\Delta}^{\pm +\Delta\Delta }=&   \mathcal{T}_{NNN\Delta}^{\pm \Delta\Delta} 
\frac{-9\sqrt{3}\ga^4}{160\fpi^4}  
(q^i q^j S_2^{i j \dagger}) \biggl[A_2^{\Delta\Delta \pm}\!\! - 7 A_3^{\Delta\Delta \pm}\!\! + 7 B_4^{\Delta\Delta \pm} \nonumber\\*
&+q^2 (B_2^{\Delta\Delta \pm}\!\! - 2 B_3^{\Delta\Delta \pm}\!\! + C_4^{\Delta\Delta \pm})\biggr]	
\label{eq:potentials_NNND}  &
\\\nonumber\\
\shortintertext{\underline{III) N$\Delta\rightarrow$N$\Delta$:}}
V_{N\Delta N\Delta}^{+ -NN }=&   \mathcal{T}_{N\Delta N\Delta}^{+ NN}
\frac{3\ga^4}{64\fpi^4}   	
 A_2^{NN-} (\vec q \! \cdot \! \vec{\sigma }_1 \; \vec q \! \cdot \! \vec{\Sigma }_2-q^2 \vec{ \sigma }_1 \! \cdot \! \vec{ \Sigma }_2) \nonumber\\ 
V_{N\Delta N\Delta}^{- +NN }=&   \mathcal{T}_{N\Delta N\Delta}^{- NN}
\frac{3\ga^4}{32\fpi^4}
\biggl\lbrace - (q^i q^j \Sigma _2^{i j}) \Bigl[A_2^{NN -}\!\! - 7 A_3^{NN -}\!\! + 7 B_4^{NN -} \nonumber\\*
& +q^2 (B_2^{NN -}\!\! - 2 B_3^{NN -}\!\!+ C_4^{NN -})\Bigr] + q^2 \Bigl[A_2^{NN -}\!\! - 10 A_3^{NN -}\!\! + 10 B_4^{NN -} \nonumber\\*
&+q^2 (B_2^{NN -}\!\! - 2 B_3^{NN -}\!\! + C_4^{NN -})\Bigr] +15 A_4^{NN -}\biggr\rbrace 	
\nonumber\\ 
V_{N\Delta N\Delta}^{+ -N\Delta }=&   \mathcal{T}_{N\Delta N\Delta}^{+ N\Delta} 
\frac{\ga^4}{400\fpi^4}  	
A_2^{N\Delta -} (\vec q \! \cdot \! \vec{\sigma }_1 \; \vec q \! \cdot \! \vec{\Sigma }_2 - q^2 \vec{ \sigma }_1 \! \cdot \! \vec{ \Sigma }_2) \nonumber\\ 
V_{N\Delta N\Delta}^{- +N\Delta }=&   \mathcal{T}_{N\Delta N\Delta}^{- N\Delta} 
\frac{\ga^4}{400\fpi^4}	
\biggl\lbrace 4 q^i q^j \Sigma _2^{i j} \Bigl[A_2^{N\Delta -}\!\! - 7 A_3^{N\Delta -}\!\! + 7 B_4^{N\Delta -} \nonumber\\*
& +q^2 (B_2^{N\Delta -}\!\! - 2 B_3^{N\Delta -}\!\! + C_4^{N\Delta -})\Bigr] + 5 q^2 \Bigl[A_2^{N\Delta -}\!\! - 10 A_3^{N\Delta -}\!\! + 10 B_4^{N\Delta -} \nonumber\\*
&+q^2 (B_2^{N\Delta -}\!\! - 2 B_3^{N\Delta -}\!\! + C_4^{N\Delta -})\Bigr] + 75 A_4^{N\Delta -}\biggr\rbrace	 \nonumber\\ 
V_{N\Delta N\Delta}^{+ -\Delta N }=&   \mathcal{T}_{N\Delta N\Delta}^{+ \Delta N}  
\frac{9\ga^4}{128\fpi^4} 
 A_2^{N\Delta -} (q^2 \vec{ \sigma }_1 \! \cdot \! \vec{ \Sigma }_2 - \vec q \! \cdot \! \vec{\sigma }_1 \; \vec q \! \cdot \! \vec{\Sigma }_2)	\nonumber\\ 
V_{N\Delta N\Delta}^{- +\Delta N }=&   \mathcal{T}_{N\Delta N\Delta}^{- \Delta N}  
\frac{9\ga^4}{32\fpi^4} 
\biggl\lbrace - (q^i q^j \Sigma _2^{i j}) \Bigl[A_2^{N\Delta -}\!\! - 7 A_3^{N\Delta -}\!\! + 7 B_4^{N\Delta -} \nonumber\\*
& +q^2 (B_2^{N\Delta -}\!\! - 2 B_3^{N\Delta -}\!\!+ C_4^{N\Delta -})\Bigr] + q^2 \Bigl[A_2^{N\Delta -}\!\! - 10 A_3^{N\Delta -}\!\! + 10 B_4^{N\Delta -} \nonumber\\*
&+q^2 (B_2^{N\Delta -}\!\! - 2 B_3^{N\Delta -}\!\! + C_4^{N\Delta -})\Bigr] +15 A_4^{N\Delta -}\biggr\rbrace 
\nonumber\\ 
V_{N\Delta N\Delta}^{\pm -\Delta\Delta }=&   \mathcal{T}_{N\Delta N\Delta}^{\pm \Delta\Delta}  
\frac{3\ga^4}{800\fpi^4} 
 A_2^{\Delta\Delta \mp} (q^2 \vec{ \sigma }_1 \! \cdot \! \vec{ \Sigma }_2 - \vec q \! \cdot \! \vec{\sigma }_1 \; \vec q \! \cdot \! \vec{\Sigma }_2)	\nonumber\\ 
V_{N\Delta N\Delta}^{\pm +\Delta\Delta }=&   \mathcal{T}_{N\Delta N\Delta}^{\pm \Delta\Delta}
\frac{3\ga^4}{400\fpi^4}  
\biggl\lbrace 4 q^i q^j \Sigma _2^{i j} \Bigl[A_2^{\Delta\Delta \pm}\!\! - 7 A_3^{\Delta\Delta \pm}\!\! + 7 B_4^{\Delta\Delta \pm} \nonumber\\*
& +q^2 (B_2^{\Delta\Delta \pm}\!\! - 2 B_3^{\Delta\Delta \pm}\!\! + C_4^{\Delta\Delta \pm})\Bigr] + 5 q^2 \Bigl[A_2^{\Delta\Delta \pm}\!\! - 10 A_3^{\Delta\Delta \pm}\!\! + 10 B_4^{\Delta\Delta \pm} \nonumber\\*
&+q^2 (B_2^{\Delta\Delta \pm}\!\! - 2 B_3^{\Delta\Delta \pm}\!\! + C_4^{\Delta\Delta \pm})\Bigr] + 75 A_4^{\Delta\Delta \pm}\biggr\rbrace
\label{eq:potentials_NDND} &
\\\nonumber\\
\shortintertext{\underline{IV) $\Delta$N$\rightarrow$N$\Delta$:}}
V_{\Delta NN\Delta}^{+ -NN }=&   \mathcal{T}_{\Delta NN\Delta}^{+ NN}
\frac{9\ga^4}{128\fpi^4}  
A_2^{NN-} (\vec q \! \cdot \! \vec{S}_1 \; \vec q \! \cdot \! \vec{S}_2^{\dagger}-q^2 \vec{ S}_1 \! \cdot \! \vec{ S}_2^{\dagger})	\nonumber\\ 
V_{\Delta NN\Delta}^{- +NN }=&   \mathcal{T}_{\Delta NN\Delta}^{- NN}
\frac{27\ga^4}{64\fpi^4}    
\biggl[(A_2^{NN -}\!\! - 4 A_3^{NN -}\!\! + 4 B_4^{NN -}) (q^i q^j S_1^{i k} S_2^{j k \dagger})\nonumber\\*
&+2 A_4^{NN -} (S_1^{i j} S_2^{i j \dagger}) + (B_2^{NN -}\!\! - 2 B_3^{NN -}\!\! + C_4^{NN -}) (q^i q^j q^k q^l S_1^{i j} S_2^{k l \dagger})\biggr]		\nonumber\\ 
V_{\Delta NN\Delta}^{+ -N\Delta }=&   \mathcal{T}_{\Delta NN\Delta}^{+ N\Delta}
\frac{9\ga^4}{128\fpi^4}   
A_2^{N\Delta -} (q^2 \vec{ S}_1 \! \cdot \! \vec{ S}_2^{\dagger} - \vec q \! \cdot \! \vec{S}_1 \; \vec q \! \cdot \! \vec{S}_2^{\dagger})	\nonumber\\ 
V_{\Delta NN\Delta}^{- +N\Delta }=&   \mathcal{T}_{\Delta NN\Delta}^{- N\Delta}
\frac{27\ga^4}{320\fpi^4}   
\biggl[(A_2^{N\Delta -}\!\! - 4 A_3^{N\Delta -}\!\! + 4 B_4^{N\Delta -}) (q^i q^j S_1^{i k} S_2^{j k \dagger})\nonumber\\*
&+2 A_4^{N\Delta -} (S_1^{i j} S_2^{i j \dagger}) + (B_2^{N\Delta -}\!\! - 2 B_3^{N\Delta -}\!\! + C_4^{N\Delta -}) (q^i q^j q^k q^l S_1^{i j} S_2^{k l \dagger})\biggr]		\nonumber\\ 
V_{\Delta NN\Delta}^{+ -\Delta N }=&   \mathcal{T}_{\Delta NN\Delta}^{+ \Delta N}
\frac{9\ga^4}{128\fpi^4}
A_2^{N\Delta -} (q^2 \vec{ S}_1 \! \cdot \! \vec{ S}_2^{\dagger} - \vec q \! \cdot \! \vec{S}_1 \; \vec q \! \cdot \! \vec{S}_2^{\dagger}) 	\nonumber\\ 
V_{\Delta NN\Delta}^{- +\Delta N }=&   \mathcal{T}_{\Delta NN\Delta}^{- \Delta N}
\frac{27\ga^4}{320\fpi^4}
\biggl[(A_2^{N\Delta -}\!\! - 4 A_3^{N\Delta -}\!\! + 4 B_4^{N\Delta -}) (q^i q^j S_1^{i k} S_2^{j k \dagger})\nonumber\\*
&+2 A_4^{N\Delta -} (S_1^{i j} S_2^{i j \dagger}) + (B_2^{N\Delta -}\!\! - 2 B_3^{N\Delta -}\!\! + C_4^{N\Delta -}) (q^i q^j q^k q^l S_1^{i j} S_2^{k l \dagger})\biggr]		\nonumber\\ 
V_{\Delta NN\Delta}^{\pm -\Delta\Delta }=&   \mathcal{T}_{\Delta NN\Delta}^{\pm \Delta\Delta}
\frac{9\ga^4}{128\fpi^4}
A_2^{\Delta\Delta \mp} (\vec q \! \cdot \! \vec{S}_1 \; \vec q \! \cdot \! \vec{S}_2^{\dagger}-q^2 \vec{ S}_1 \! \cdot \! \vec{ S}_2^{\dagger})	\nonumber\\
V_{\Delta NN\Delta}^{\pm +\Delta\Delta }=&   \mathcal{T}_{\Delta NN\Delta}^{\pm \Delta\Delta} 
\frac{27\ga^4}{1600\fpi^4}
\biggl[(A_2^{\Delta\Delta \pm}\!\! - 4 A_3^{\Delta\Delta \pm}\!\! + 4 B_4^{\Delta\Delta \pm}) (q^i q^j S_1^{i k} S_2^{j k \dagger})\nonumber\\*
&+2 A_4^{\Delta\Delta \pm} (S_1^{i j} S_2^{i j \dagger}) + (B_2^{\Delta\Delta \pm}\!\! - 2 B_3^{\Delta\Delta \pm}\!\! + C_4^{\Delta\Delta \pm}) (q^i q^j q^k q^l S_1^{i j} S_2^{k l \dagger})\biggr]	
\label{eq:potentials_DNND} &
\\\nonumber\\
\shortintertext{\underline{V) NN$\rightarrow \Delta\Delta$:}}
V_{NN\Delta\Delta}^{+ -NN }=&   \mathcal{T}_{NN\Delta\Delta}^{+ NN}
\frac{9\ga^4}{128\fpi^4}    
A_2^{NN-} (\vec q \! \cdot \! \vec{S}_1^{\dagger} \; \vec q \! \cdot \! \vec{S}_2^{\dagger}-q^2 \vec{ S}_1^{\dagger} \! \cdot \! \vec{S}_2^{\dagger})	\nonumber\\ 
V_{NN\Delta\Delta}^{- +NN }=&   \mathcal{T}_{NN\Delta\Delta}^{- NN}
\frac{27\ga^4}{64\fpi^4}    
\biggl[ (A_2^{NN -}\!\! - 4 A_3^{NN -}\!\! + 4 B_4^{NN -}) (q^i q^j S_1^{i k \dagger} S_2^{j k \dagger})\nonumber\\*
&+2 A_4^{NN -} (S_1^{i j \dagger} S_2^{i j \dagger})+(B_2^{NN -}\!\! - 2 B_3^{NN -}\!\! + C_4^{NN -}) (q^i q^j q^k q^l S_1^{i j \dagger} S_2^{k l \dagger})\biggr]	\nonumber\\ 
V_{NN\Delta\Delta}^{+ -N\Delta }=&   \mathcal{T}_{NN\Delta\Delta}^{+ N\Delta}\!
\frac{9\ga^4}{128\fpi^4}  
A_2^{N\Delta -} (q^2 \vec{ S}_1^{\dagger} \! \cdot \! \vec{S}_2^{\dagger} -\vec q \! \cdot \! \vec{S}_1^{\dagger} \; \vec q \! \cdot \! \vec{S}_2^{\dagger})	\nonumber\\
V_{NN\Delta\Delta}^{- +N\Delta }=&   \mathcal{T}_{NN\Delta\Delta}^{- N\Delta} \!
\frac{27\ga^4}{320\fpi^4}  
\biggl[ (A_2^{N\Delta -}\!\! - 4 A_3^{N\Delta -}\!\! + 4 B_4^{N\Delta -}) (q^i q^j S_1^{i k \dagger} S_2^{j k \dagger})\nonumber\\*
&+2 A_4^{N\Delta -} (S_1^{i j \dagger} S_2^{i j \dagger})+(B_2^{N\Delta -}\!\! - 2 B_3^{N\Delta -}\!\! + C_4^{N\Delta -}) (q^i q^j q^k q^l S_1^{i j \dagger} S_2^{k l \dagger})\biggr] \nonumber\\
V_{NN\Delta\Delta}^{\pm -\Delta\Delta }=&   \mathcal{T}_{NN\Delta\Delta}^{\pm \Delta\Delta} 
\frac{9\ga^4}{128\fpi^4}  
A_2^{\Delta\Delta -} (\vec q \! \cdot \! \vec{S}_1^{\dagger} \; \vec q \! \cdot \! \vec{S}_2^{\dagger}-q^2 \vec{ S}_1^{\dagger} \! \cdot \! \vec{S}_2^{\dagger})	\nonumber\\ 
V_{NN\Delta\Delta}^{\pm +\Delta\Delta }=&   \mathcal{T}_{NN\Delta\Delta}^{\pm \Delta\Delta}
\frac{27\ga^4}{1600\fpi^4}
\biggl[ (A_2^{\Delta\Delta \pm}\!\! - 4 A_3^{\Delta\Delta \pm}\!\! + 4 B_4^{\Delta\Delta \pm}) (q^i q^j S_1^{i k \dagger} S_2^{j k \dagger})\nonumber\\*
&+2 A_4^{\Delta\Delta \pm} (S_1^{i j \dagger} S_2^{i j \dagger})+(B_2^{\Delta\Delta \pm}\!\! - 2 B_3^{\Delta\Delta \pm}\!\! + C_4^{\Delta\Delta \pm}) (q^i q^j q^k q^l S_1^{i j \dagger} S_2^{k l \dagger})\biggr]
\label{eq:potentials_NNDD} &
\\\nonumber\\
\shortintertext{\underline{VI) N$\Delta \rightarrow \Delta\Delta$:}}
V_{N\Delta\Delta\Delta}^{+ -NN }=&   \mathcal{T}_{N\Delta\Delta\Delta}^{+ NN}
\frac{9\ga^4}{128\sqrt{2}\fpi^4} 
A_2^{NN-} (q^2 \vec{ S}_1^{\dagger} \! \cdot \! \vec{\Sigma}_2 -\vec q \! \cdot \! \vec{S}_1^{\dagger} \; \vec q \! \cdot \! \vec{\Sigma}_2) \nonumber\\ 
V_{N\Delta\Delta\Delta}^{- +NN }=&   \mathcal{T}_{N\Delta\Delta\Delta}^{- NN}
\frac{9\sqrt{3}\ga^4}{64\fpi^4}  	 
\biggl\lbrace -\Bigl[A_2^{NN-}\!\! - 7 A_3^{NN-}\!\! + 7B_4^{NN-} \nonumber\\*
&+ q^2 (B_2^{NN-}\!\! - 2 B_3^{NN-}\!\! + C_4^{NN-})\Bigr] (q^i q^j S_1^{ij \dagger})\nonumber\\* 
&+ (A_2^{NN-}\!\! - 4 A_3^{NN-}\!\! + 4 B_4^{NN-}) (q^i q^j S_1^{i k \dagger} \Sigma _2^{j k}) +2 A_4^{NN-} (S_1^{i j \dagger} \Sigma _2^{i j})\nonumber\\*
&+ (B_2^{NN-}\!\! - 2 B_3^{NN-}\!\! + C_4^{NN-}) (q^i q^j q^k q^l S_1^{i j \dagger} \Sigma _2^{k l})\biggr\rbrace
\nonumber\\ 
V_{N\Delta\Delta\Delta}^{+ -N\Delta }=&   \mathcal{T}_{N\Delta\Delta\Delta}^{+ N\Delta}
\frac{3\ga^4}{800\sqrt{2}\fpi^4} 
A_2^{N\Delta -} (q^2 \vec{ S}_1^{\dagger} \! \cdot \! \vec{\Sigma}_2 -\vec q \! \cdot \! \vec{S}_1^{\dagger} \; \vec q \! \cdot \! \vec{\Sigma}_2) 	\nonumber\\ 
V_{N\Delta\Delta\Delta}^{- +N\Delta }=&   \mathcal{T}_{N\Delta\Delta\Delta}^{- N\Delta}
\frac{-3\sqrt{3}\ga^4}{800\fpi^4} 	
\biggl\lbrace 5 \Bigl[ A_2^{N\Delta -}-7 A_3^{N\Delta -} +7 B_4^{N\Delta -} \nonumber\\*
&+q^2 (B_2^{N\Delta -}-2 B_3^{N\Delta -}+C_4^{N\Delta -})\Bigr] (q^i q^j S_1^{i j \dagger}) \nonumber\\*
&+4 (A_2^{N\Delta -}-4 A_3^{N\Delta -}+4 B_4^{N\Delta -}) (q^i q^j S_1^{i k \dagger} \Sigma _2^{j k})+8 A_4^{N\Delta -} 	(S_1^{i j \dagger} \Sigma _2^{i j})\nonumber\\*
&+4 (B_2^{N\Delta -}-2 B_3^{N\Delta -}+C_4^{N\Delta -}) (q^i q^j q^k q^l S_1^{i j \dagger} \Sigma _2^{k l})\biggr\rbrace		\nonumber\\ 
V_{N\Delta\Delta\Delta}^{+ -\Delta N }=&   \mathcal{T}_{N\Delta\Delta\Delta}^{+ \Delta N} 
\frac{9\ga^4}{128\sqrt{2}\fpi^4}   
A_2^{N\Delta -} (\vec q \! \cdot \! \vec{S}_1^{\dagger} \; \vec q \! \cdot \! \vec{\Sigma}_2-q^2 \vec{ S}_1^{\dagger} \! \cdot \! \vec{\Sigma}_2)	\nonumber\\ 
V_{N\Delta\Delta\Delta}^{- +\Delta N }=&   \mathcal{T}_{N\Delta\Delta\Delta}^{- \Delta N} 
\frac{9\sqrt{3}\ga^4}{320\fpi^4} 
\biggl\lbrace -\Bigl[A_2^{N\Delta -}\!\! - 7 A_3^{N\Delta -}\!\! + 7B_4^{N\Delta -} \nonumber\\*
&+ q^2 (B_2^{N\Delta -}\!\! - 2 B_3^{N\Delta -}\!\! + C_4^{N\Delta -})\Bigr] (q^i q^j S_1^{ij \dagger})\nonumber\\* 
&+ (A_2^{N\Delta -}\!\! - 4 A_3^{N\Delta -}\!\! + 4 B_4^{N\Delta -}) (q^i q^j S_1^{i k \dagger} \Sigma _2^{j k}) +2 A_4^{N\Delta -} (S_1^{i j \dagger} \Sigma _2^{i j})\nonumber\\*
&+ (B_2^{N\Delta -}\!\! - 2 B_3^{N\Delta -}\!\! + C_4^{N\Delta -}) (q^i q^j q^k q^l S_1^{i j \dagger} \Sigma _2^{k l})\biggr\rbrace	\nonumber\\ 
V_{N\Delta\Delta\Delta}^{\pm -\Delta\Delta }=&   \mathcal{T}_{N\Delta\Delta\Delta}^{\pm \Delta\Delta} 
\frac{3\ga^4}{800\sqrt{2}\fpi^4} 
A_2^{\Delta\Delta \mp} (\vec q \! \cdot \! \vec{S}_1^{\dagger} \; \vec q \! \cdot \! \vec{\Sigma}_2-q^2 \vec{ S}_1^{\dagger} \! \cdot \! \vec{\Sigma}_2) \nonumber\\ 
V_{N\Delta\Delta\Delta}^{\pm +\Delta\Delta }=&   \mathcal{T}_{N\Delta\Delta\Delta}^{\pm \Delta\Delta} 
\frac{-3\sqrt{3}\ga^4}{4000\fpi^4} 
\biggl\lbrace 5 \Bigl[ A_2^{\Delta\Delta \pm}-7 A_3^{\Delta\Delta \pm} +7 B_4^{\Delta\Delta \pm} \nonumber\\*
&+q^2 (B_2^{\Delta\Delta \pm}-2 B_3^{\Delta\Delta \pm}+C_4^{\Delta\Delta \pm})\Bigr] (q^i q^j S_1^{i j \dagger}) \nonumber\\*
&+4 (A_2^{\Delta\Delta \pm}-4 A_3^{\Delta\Delta \pm}+4 B_4^{\Delta\Delta \pm}) (q^i q^j S_1^{i k \dagger} \Sigma _2^{j k})+8 A_4^{\Delta\Delta \pm} 	(S_1^{i j \dagger} \Sigma _2^{i j})\nonumber\\*
&+4 (B_2^{\Delta\Delta \pm}-2 B_3^{\Delta\Delta \pm}+C_4^{\Delta\Delta \pm}) (q^i q^j q^k q^l S_1^{i j \dagger} \Sigma _2^{k l})\biggr\rbrace	
\label{eq:potentials_NDDD}  &
\\\nonumber\\
\shortintertext{\underline{VII) $\Delta\Delta \rightarrow \Delta\Delta$:}}
V_{\Delta\Delta\Delta\Delta}^{+ -NN }=&   \mathcal{T}_{\Delta\Delta\Delta\Delta}^{+ NN}
\frac{9\ga^4}{256\fpi^4}  
A_2^{NN -} (\vec q \! \cdot \! \vec{\Sigma }_1 \; \vec q \! \cdot \! \vec{\Sigma }_2-q^2 \vec{ \Sigma }_1 \! \cdot \! \vec{ \Sigma }_2)
\nonumber\\
V_{\Delta\Delta\Delta\Delta}^{- +NN }=&   \mathcal{T}_{\Delta\Delta\Delta\Delta}^{- NN}
\frac{9\ga^4}{64\fpi^4}   
\biggl\lbrace  q^i q^j q^k q^l \Sigma_{1}^{ij} \Sigma_{2}^{kl} ( B_2 -2 B_3+C_4) \nonumber\\*
&+  q^i q^j \Sigma_{1}^{ik} \Sigma_{2}^{jk} ( A_2^{NN-}\!\! - 4 A_3^{NN-}\!\! + 4 B_4^{NN-}  )  + 2 \Sigma_{1}^{ij} \Sigma_{2}^{ij} A_4^{NN-}  \nonumber\\*
&-  q^i q^j (\Sigma_{1}^{ij}+\Sigma_{2}^{ij}) \Bigl[ A_2^{NN-}\!\! - 7 A_3^{NN-}\!\! + 7 B_4^{NN-}    \nonumber\\*
&+ q^2 ( B_2^{NN-}\!\! - 2 B_3^{NN-}\!\! +  C_4^{NN-}) \Bigr] + \Bigl[ 15 A_4^{NN-}    \nonumber\\*
&+ q^2 ( A_2^{NN-}\!\! - 10 A_3^{NN-}\!\! + 10 B_4^{NN-}) + q^4 ( B_2^{NN-}\!\! - 2 B_3^{NN-}\!\! + C_4^{NN-})\Bigr]\biggr\rbrace 
\nonumber\\
V_{\Delta\Delta\Delta\Delta}^{+ -N\Delta }=&   \mathcal{T}_{\Delta\Delta\Delta\Delta}^{+ N\Delta} 
\frac{3\ga^4}{1600\fpi^4}     
A_2^{N \Delta -} (\vec q \! \cdot \! \vec{\Sigma }_1 \; \vec q \! \cdot \! \vec{\Sigma }_2-q^2 \vec{ \Sigma }_1 \! \cdot \! \vec{ \Sigma }_2)
\nonumber\\ 
V_{\Delta\Delta\Delta\Delta}^{- +N\Delta }=&   \mathcal{T}_{\Delta\Delta\Delta\Delta}^{- N\Delta} 
\frac{3\ga^4}{800\fpi^4}  
\biggl\lbrace -4 q^i q^j q^k q^l \Sigma_{1}^{ij} \Sigma_{2}^{kl} ( B_2^{N\Delta -}\!\! - 2 B_3^{N\Delta -}\!\! +C_4^{N\Delta -} ) \nonumber\\*
&+ 4 q^i q^j \Sigma_{1}^{ik} \Sigma_{2}^{jk} ( A_2^{N\Delta -}\!\! - 4 A_3^{N\Delta -}\!\! + 4 B_4^{N\Delta -} )  - 8 \Sigma_{1}^{ij} \Sigma_{2}^{ij} A_4^{N\Delta -}  \nonumber\\*
&- q^i q^j (5\Sigma_{1}^{ij}-4\Sigma_{2}^{ij}) \Bigl[ A_2^{N\Delta -} - 7 A_3^{N\Delta -} + 7 B_4^{N\Delta -}    \nonumber\\*
&+ q^2 ( B_2^{N\Delta -}\!\! - 2 B_3^{N\Delta -} +  C_4^{N\Delta -}) \Bigr] + 5 \Bigl[ 15 A_4^{N\Delta -}    \nonumber\\*
&+ q^2 ( A_2^{N\Delta -}\!\! - 10 A_3^{N\Delta -}\!\! + 10 B_4^{N\Delta -}) + q^4 ( B_2^{N\Delta -}\!\! - 2 B_3^{N\Delta -}\!\! + C_4^{N\Delta -} )\Bigr] \biggr\rbrace 
\nonumber\\
V_{\Delta\Delta\Delta\Delta}^{\pm -\Delta\Delta }=&  \mathcal{T}_{\Delta\Delta\Delta\Delta}^{\pm\Delta\Delta}
\frac{\ga^4}{10000\fpi^4}   
A_2^{\Delta \Delta \mp}( q^{i} q^{j} \Sigma_{1}^{i} \Sigma_{2}^{j} - q^2 \Sigma_{1}^{i} \Sigma_{2}^{i})	
\nonumber\\ 
V_{\Delta\Delta\Delta\Delta}^{\pm +\Delta\Delta }=&  \mathcal{T}_{\Delta\Delta\Delta\Delta}^{\pm \Delta\Delta} 
\frac{\ga^4}{10000\fpi^4}   
\biggl\lbrace 16 q^i q^j q^k q^l \Sigma_{1}^{ij} \Sigma_{2}^{kl} ( B_2^{\Delta\Delta \pm}\!\! - 2 B_3^{\Delta\Delta \pm}\!\! + C_4^{\Delta\Delta \pm}) \nonumber\\*
&+ 16 q^i q^j \Sigma_{1}^{ik} \Sigma_{2}^{jk} ( A_2^{\Delta\Delta \pm}\!\! - 4 A_3^{\Delta\Delta \pm}\!\! + 4 B_4^{\Delta\Delta \pm}  + 32 \Sigma_{1}^{ij} \Sigma_{2}^{ij} A_4^{\Delta\Delta \pm}\nonumber\\*
&+ 20 q^i q^j (\Sigma_{1}^{ij}+\Sigma_{2}^{ij}) \Bigl[ A_2^{\Delta\Delta \pm}\!\! - 7 A_3^{\Delta\Delta \pm}\!\! + 7 B_4^{\Delta\Delta \pm}  \nonumber\\*
&+ q^2 ( B_2^{\Delta\Delta \pm}\!\! - 2 B_3^{\Delta\Delta \pm}\!\! +  C_4^{\Delta\Delta \pm}) \Bigr] +25 \Bigl[ 15 A_4^{\Delta\Delta \pm}  \nonumber\\*
&+ q^2 ( A_2^{\Delta\Delta \pm}\!\! - 10 A_3^{\Delta\Delta \pm}\!\! + 10 B_4^{\Delta\Delta \pm} )+q^4 ( B_2^{\Delta\Delta \pm}\!\! - 2 B_3^{\Delta\Delta \pm}\!\! +   C_4^{\Delta\Delta \pm} )\Bigr] \biggr\rbrace
\label{eq:potentials_DDDD} &
\end{align}

\subsubsection{Triangle diagrams}
The set of contributing $2\pi$-exchange triangle diagrams is shown in lines (h) and (i) of \cref{fig:OPE_TPE_diagrams}.
The triangle diagrams have a single baryon (N or $\Delta$) in the intermediate state and the corresponding imaginary parts $\Im A_1 = \widetilde A_1$, $\Im A_2 = \widetilde  A_2$, $\Im B_2 = \widetilde B_2$ defined in \cref{eq:coefficients_1_l,eq:coefficients_2_l_a,eq:coefficients_2_l_b} are given by the following expressions
\begin{align}
\Im A_{1}^{N} &= \frac{w}{16 \pi \mu} \; , \nonumber\\*
\Im A_{2}^{N} &= \frac{w^3}{96 \pi \mu} \; , \nonumber\\*
\Im B_{2}^{N} &= \frac{w}{24 \pi \mu^3} (\mu^2-\mpi^2) \; , \\\nonumber\\
\Im A_{1}^{\Delta} &= \frac{1}{48 \pi \mu} \biggl[\frac{w}{2}-\Delta \arctan \frac{w}{2\Delta}\biggr] \; , \nonumber\\*
\Im A_{2}^{\Delta} &= \frac{1}{192 \pi \mu} \biggl[12\Delta^2 w + 2 w^3 - 6 \Delta (4\Delta^2 + w^2) \arctan \frac{w}{2\Delta} \biggr] \; , \nonumber\\*
\Im B_{2}^{\Delta} &= \frac{1}{96 \pi \mu^3} \biggl[6 \Delta^2 w + 4 w^3  - 3 \Delta (4\Delta^2 -4 \mpi^2 + 3 \mu^2)\arctan \frac{w}{2\Delta}\biggr] \; ,
\end{align}
with $w=\sqrt{\mu^2-4 \mpi^2}$.
The $2\pi$-exchange potentials from the triangle diagrams with a nucleon intermediate state read
\begin{align}
V_{NNNN}^{\diagramtriangle[0.5]} =& \frac{\ga^2}{8 \fpi^4} (4I-3)  \biggl[ q^2 A_{1}^{ N} - 3 A_{2}^{ N} - q^2 B_{2}^{ N}  \biggr] \; , \nonumber\\
V_{NNNN}^{\diagramtrianglemirrored[0.5]}=& \frac{\ga^2}{8 \fpi^4} (4I-3) \biggl[ q^2 A_{1}^{ N} - 3 A_{2}^{ N} - q^2 B_{2}^{ N}  \biggr] \; , \nonumber\\ 
V_{NNN\Delta}^{\diagramtriangle[0.5]}=& \frac{3\ga^2}{8 \sqrt{2} \fpi^4} I   S_{2}^{ij\dagger} \biggl[ q^{i} q^{j} A_{1}^{ N} - \delta^{ij} A_{2}^{ N} - q^{i} q^{j} B_{2}^{ N}  \biggr] \; , \nonumber\\ 
V_{N\Delta N\Delta}^{\diagramtriangle[0.5]} =& \frac{-15\sqrt{3}\ga^2}{32\sqrt{2} \fpi^4}  I  S_{2}^{ij\dagger} \biggl[ q^{i} q^{j} A_{1}^{ N} - \delta^{ij} A_{2}^{ N} - q^{i} q^{j} B_{2}^{ N}  \biggr] \; , \nonumber\\
V_{N\Delta N\Delta}^{\diagramtrianglemirrored[0.5]}=& \frac{5\ga^2}{8 \fpi^4}  I  \biggl[ q^2 A_{1}^{ N} - 3 A_{2}^{ N} - q^2 B_{2}^{ N}  \biggr] \; , \nonumber\\
V_{N\Delta\Delta\Delta}^{\diagramtrianglemirrored[0.5]}=& \frac{-3\sqrt{5}\ga^2}{16 \fpi^4}   I  S_{1}^{ij\dagger} \biggl[ q^{i} q^{j} A_{1}^{ N} - \delta^{ij} A_{2}^{ N} - q^{i} q^{j} B_{2}^{ N}  \biggr] \; , \nonumber\\
V_{\Delta\Delta\Delta\Delta}^{\diagramtriangle[0.5]} =& \frac{3\sqrt{3}\ga^2}{32 \sqrt{2} \fpi^4} (15 - 4I) S_{2}^{ij\dagger} \biggl[ q^{i} q^{j} A_{1}^{ N} - \delta^{ij} A_{2}^{ N} - q^{i} q^{j} B_{2}^{ N}  \biggr] \; , \nonumber\\
V_{\Delta\Delta\Delta\Delta}^{\diagramtrianglemirrored[0.5]}=& \frac{3\sqrt{3}\ga^2}{32 \sqrt{2} \fpi^4}  (15 - 4I) S_{1}^{ij\dagger} \biggl[ q^{i} q^{j} A_{1}^{ N} - \delta^{ij} A_{2}^{ N} - q^{i} q^{j} B_{2}^{ N}  \biggr] \; ,  
\end{align}

and those from triangle diagrams with a $\Delta$ intermediate state take the form

\begin{align}
V_{NNNN}^{\diagramtriangleD[0.5]}=& \frac{3\sqrt{3}\ga^2}{16\sqrt{2} \fpi^4} (4I - 3) S_{2}^{ij\dagger} \biggl[ q^{i} q^{j} A_{1}^{ \Delta} - \delta^{ij} A_{2}^{ \Delta} - q^{i} q^{j} B_{2}^{ \Delta}  \biggr] \; , \nonumber\\
V_{NNNN}^{\diagramtriangleDmirrored[0.5]}=& \frac{3\sqrt{3}\ga^2}{16\sqrt{2} \fpi^4}  (4I - 3) S_{1}^{ij\dagger} \biggl[ q^{i} q^{j} A_{1}^{ \Delta} - \delta^{ij} A_{2}^{ \Delta} - q^{i} q^{j} B_{2}^{ \Delta}  \biggr] \; , \nonumber\\
V_{NNN\Delta}^{\diagramtriangleD[0.5]}=& \frac{-3\ga^2}{8 \sqrt{2} \fpi^4} I  S_{2}^{ij\dagger}  \biggl[ q^{i} q^{j} A_{1}^{ \Delta} - \delta^{ij} A_{2}^{ \Delta} - q^{i} q^{j} B_{2}^{ \Delta}  \biggr] \; , \nonumber\\
V_{N\Delta N\Delta}^{\diagramtriangleD[0.5]}=& \frac{-\ga^2}{40 \fpi^4} I (5\delta_{2}^{ij} + 4 \Sigma_{2}^{ij}) \biggl[ q^{i} q^{j} A_{1}^{ \Delta} - \delta^{ij} A_{2}^{ \Delta} - q^{i} q^{j} B_{2}^{ \Delta}  \biggr] \; , \nonumber\\
V_{N\Delta N\Delta}^{\diagramtriangleDmirrored[0.5]}=& \frac{-15\sqrt{3}\ga^2}{16\sqrt{2} \fpi^4}  I S_{1}^{ij\dagger} \biggl[ q^{i} q^{j} A_{1}^{ \Delta} - \delta^{ij} A_{2}^{ \Delta} - q^{i} q^{j} B_{2}^{ \Delta}  \biggr] \; , \nonumber\\
V_{N\Delta\Delta\Delta}^{\diagramtriangleDmirrored[0.5]}=& \frac{3\sqrt{5}\ga^2}{16 \fpi^4} I  S_{1}^{ij\dagger} \biggl[ q^{i} q^{j} A_{1}^{ \Delta} - \delta^{ij} A_{2}^{ \Delta} - q^{i} q^{j} B_{2}^{ \Delta}  \biggr] \; , \nonumber\\
V_{\Delta\Delta\Delta\Delta}^{\diagramtriangleD[0.5]}=& \frac{\ga^2}{200 \fpi^4} (4I-15) (5\delta_{2}^{ij} + 4 \Sigma_{2}^{ij}) \biggl[ q^{i} q^{j} A_{1}^{ \Delta} - \delta^{ij} A_{2}^{ \Delta} - q^{i} q^{j} B_{2}^{ \Delta}  \biggr] \; , \nonumber\\
V_{\Delta\Delta\Delta\Delta}^{\diagramtriangleDmirrored[0.5]}=& \frac{\ga^2}{200 \fpi^4}  (4I-15) (5\delta_{1}^{ij} + 4 \Sigma_{1}^{ij}) \biggl[ q^{i} q^{j} A_{1}^{ \Delta} - \delta^{ij} A_{2}^{ \Delta} - q^{i} q^{j} B_{2}^{ \Delta}  \biggr]  \; ,
\end{align}
where $I=0,1$ is the total isospin. Note that the left and right triangle diagram have been carefully distinguished, although they give in some cases identical results.

\subsubsection{Bubble diagrams}
The $2\pi$-exchange bubble diagrams with identical initial and final states are shown in line (j) of \cref{fig:OPE_TPE_diagrams}. The imaginary parts for the three non-vanishing potentials read
\begin{align}
\Im V_{NNNN}^{\diagramfootball[0.5]} =&  \frac{w^3}{768 \fpi^4 \pi \mu} (3-4I) \; ,  \nonumber\\
\Im V_{N\Delta N\Delta}^{\diagramfootball[0.5]} =& \frac{5w^3}{768 \fpi^4 \pi \mu}  I \; , \nonumber\\
\Im V_{\Delta\Delta\Delta\Delta}^{\diagramfootball[0.5]} =&  \frac{w^3}{768 \fpi^4 \pi \mu} (15-4I) \; .  
\end{align}

\subsection{Regularization and partial wave decomposition}
\label{subsec:regularization}
The potentials for the coupled NN-, N$\Delta$-, $\Delta$N-, $\Delta\Delta$-channels derived in chiral effective field theory need to be regularized in order to cut off unphysical high-momentum components.
We employ the local regulator of Ref.~\cite{Reinert2017Semilocalmomentumspace}, which follows an earlier approach in Ref.~\cite{Rijken1991Softtwopion}. The regularization is implemented by softening the ultraviolet behavior of pion propagators, which yields for the one-pion exchange potentials in \cref{eq:OPE_NNNN,eq:OPE_NNND,eq:OPE_NDND,eq:OPE_DNND,eq:OPE_NNDD,eq:OPE_NDDD,eq:OPE_DDDD} the replacement
\begin{align}
\frac{1}{\mpi^2+q^2} ~~\rightarrow ~~ \frac{ \exp \left[- (q^2 + \mpi^2)/\Lambda^2 \right]}{\mpi^2+q^2} \; .
\label{eq:regulator_1pi}
\end{align}
As the two-pion exchange potentials are given as dispersion integrals over their imaginary parts, this method leads to a regularization already at the level of the spectral representation by replacing $V(q)$ with $V_{\Lambda}(q)$,
\begin{align}
V(q) =  \frac{2}{\pi} \int_{2\mpi}^{\infty}\limits \d \mu \frac{\mu \Im V(\iu \mu)}{\mu^2 +q^2} ~~ \rightarrow ~~
V_{\Lambda}(q) = \e^{-\frac{q^2}{2\Lambda^2}} \frac{2}{\pi} \int_{2\mpi}^{\infty}\limits \d \mu \frac{\mu \Im V(\iu \mu)}{\mu^2 +q^2} \e^{-\frac{\mu^2}{2\Lambda^2}} \;.
\label{eq:regulator_2pi}
\end{align}
This regularization is applied to all functions $A_i$, $B_i$, $C_i$, which build up (through their analytically calculated imaginary parts) the potentials from box and triangle diagrams, as well as to the potential from the bubble diagrams.
Note that this two-pion exchange regulator function provides a cutoff for the spectral integral over $\mu$ and the resulting $q$-dependent potential at the same time, such that no subtractions are necessary.
In the applications, the cutoff parameter $\Lambda$ in \cref{eq:regulator_1pi,eq:regulator_2pi} will be chosen in the range $450\dots\SI{700}{MeV}$.

In the next step one has to perform a partial wave decomposition of the (transition) potentials.
We follow the convenient and clear method of Ref.~\cite{Golak2010newwayto}, where the matrix element in spin-space is first given by a four-fold angular integral.
Due to its independence of $m_j$ (total angular momentum projection) one can average over this quantum number and ends up with a single integral over the cosine of a polar angle $z$.
To be specific the initial and final momenta $\vec{p}$ and $\vec{p}^{\;\prime}$ are chosen as
\begin{align}
\vec{p}=(0,0,p)  ,\quad \vec{p}^{\;\prime}=(p^{\prime}\sqrt{1-z^2},0,p^{\prime} z) \; , 
\end{align}
and momentum transfer is $\vec{q}=\vec{p}^{\;\prime}\!-\vec{p}$.
Putting all pieces together a transition matrix element of the potential $V(\vec q \;)$ (including two-body spin operators) in the $\ket{lsj}$-basis reads  \cite{Golak2010newwayto} 
\begin{align}
H(l^\prime , s^\prime , l, s, j)=&\, \frac{8\pi^2 }{2j+1} \sum\limits_{m_j=-j}^{j} \sum\limits_{m_l^{\prime}=-l^{\prime}}^{l^{\prime}}  C(l^{\prime},s^\prime,j;m_l^{\prime},m_j-m_l^{\prime},m_j) \sum\limits_{m_l=-l}^{l}  C(l,s,j;m_l,m_j-m_l,m_j) \nonumber\\*
& \times  \int_{-1}^{1} \d z \;  Y_{l^{\prime} m_{l}^{\prime}} (\arccos z, 0) Y_{l m_{l}}^{*} (0, 0)    \melement{s^\prime, m_{j}-m_l^{\prime}}{V(\vec{q}\;)}{s, m_{j}-m_l} \; .
\end{align}
Here, $j$ is the (conserved) total angular momentum quantum number, while $l$ and $l^\prime$ are orbital angular momentum quantum numbers, which can differ by 0, 2, 4, 6 units. Furthermore, $s$, $s^\prime \in \lbrace 0,1,2,3 \rbrace$ are quantum numbers related to the total two-baryon spin, which can differ by 0, 1, 2 units.  The symbol $C(l,s,j;m_l,m_s,m_j)$ denotes the conventional Clebsch-Gordan coefficients, subject to the constraint $m_s+m_l=m_j$, and $Y_{l m_{l}} (\arccos z, 0)$ are the spherical harmonics at vanishing azimuthal angle.
The remaining transition matrix elements in spin-space $\melement{s^\prime, m_{j}-m_l^{\prime}}{V(\vec{q}\;)}{s, m_{j}-m_{l}}$ can be calculated directly with the help of the well-known coupled spin-multiplet states $\ket{0,0}$, $\ket{1, m_s}$, $\ket{2, m_s}$ and $\ket{3, m_s}$ represented by products of one-body spin states. 

\subsection{Contact potentials}
\label{subsec:contact}
\noindent
At leading order baryon-baryon contact interactions are momentum independent and for the coupled (NN, N$\Delta$, $\Delta$N, $\Delta\Delta$) channels these contact potentials take the same form as given recently in Ref.~\cite{Haidenbauer2017Scatteringdecupletbaryons} 
\begin{align}
V_{ct,NNNN}^{(0)} &= C_S+C_T \; \vec \sigma_1 \cdot \vec \sigma_2 \; ,  \nonumber\\ 
V_{ct,NNN\Delta}^{(0)} &= C_{2,NNN\Delta}^{(0)} \; \vec \sigma_1 \cdot \vec S_2^{\dagger} \; , \nonumber\\
V_{ct,N\Delta N\Delta}^{(0)} &= V_{ct,\Delta N N\Delta}^{(0)} = C_{1,N\Delta N\Delta}^{(0)} + C_{2,N\Delta N\Delta}^{(0)} \; \vec \sigma_1 \cdot \vec \Sigma_2 \; , \nonumber\\
V_{ct,NN\Delta \Delta}^{(0)} &= C_{2,NN\Delta \Delta}^{(0)}  \; \vec S_1^{\dagger} \cdot \vec S_2^{\dagger}+ C_{3,NN\Delta \Delta}^{(0)} \; S_1^{ij\dagger}  S_2^{ij\dagger} \; , \nonumber\\
V_{ct,N\Delta \Delta \Delta}^{(0)} &= C_{2,N\Delta \Delta \Delta}^{(0)}  \; \vec S_1^{\dagger} \cdot \vec \Sigma_2+ C_{3,N\Delta \Delta \Delta}^{(0)} \; S_1^{ij\dagger}  \Sigma_2^{ij} \; , \nonumber\\
V_{ct,\Delta \Delta \Delta \Delta}^{(0)} &= C_{1,\Delta \Delta \Delta \Delta}^{(0)} + C_{2,\Delta \Delta \Delta \Delta}^{(0)}  \; \vec \Sigma_1 \cdot \vec \Sigma_2 + C_{3,\Delta \Delta \Delta \Delta}^{(0)} \; \Sigma_1^{ij}  \Sigma_2^{ij} + C_{4,\Delta \Delta \Delta \Delta}^{(0)} \; \Sigma_1^{ijk}  \Sigma_2^{ijk} \; .
\label{eq:contact_potential_lo}
\end{align}
The spin (transition) matrices $\vec \sigma$, $\vec S$, $\vec \Sigma$ and their combinations with multiple indices are defined in \cref{sec:spin_and_isospin_matrices}. One should note, that the potential $V_{ct,NNN\Delta}^{(0)}$ does not contribute due to restrictions imposed by the Pauli exclusion principle, which requires the initial NN state to be either ($s=0$, $I=1$) or ($s=1$, $I=0$) for $l=0$, whereas the final N$\Delta$ state has spin $s=1,\, 2$ and isospin $I=1,2$.
We remark that the contact potentials $V_{ct,N\Delta N\Delta}^{(0)}$ and $V_{ct,N\Delta \Delta\Delta}^{(0)}$ are of relevance only for total isospin $I=1$. For the other contact potentials $V_{ct,NNNN}^{(0)}$, $V_{ct,N \Delta N \Delta}^{(0)}$ and $V_{ct,\Delta \Delta \Delta \Delta}^{(0)}$, which contribute at total isospin $I=0,1$, the Pauli exclusion principle prohibits a doubling of low-energy constants.

At next-to-leading order baryon-baryon contact potentials depend quadratically on the momenta.
Using the definitions $\vec q = \vec p^{\;\prime}- \vec p$ and $\vec k = \frac{1}{2} ( \vec p^{\;\prime} + \vec p\;)$ for initial and final center-of-mass momenta $\vec p$ and $\vec p^{\;\prime}$, the purely nucleonic contact potential at next-to-leading order reads \cite{Epelbaum2005Twonucleonsystem}
\begin{align}
V_{ct,NNNN}^{(2)} =& \; C_{1}^{(2)} q^2 + C_{2}^{(2)} k^2 
+ \left(C_{3}^{(2)} q^2 + C_{4}^{(2)} k^2\right)  \; \vec \sigma_1 \cdot \vec \sigma_2 \nonumber\\*
&+ C_{5}^{(2)} \iu \left( \vec \sigma_{1} + \vec \sigma_{2} \right) \cdot (\vec{q}\times \vec{k})  
+ C_{6}^{(2)} \; \vec \sigma_1 \cdot \vec{q} \,\, \vec \sigma_2 \cdot \vec{q}
+ C_{7}^{(2)} \; \vec \sigma_1 \cdot \vec{k} \,\, \vec \sigma_2 \cdot \vec{k} \; .
\label{eq:contact_potential_nlo_NNNN}
\end{align}
Note that this contact potential contributes only to $S$- and $P$-waves of elastic nucleon-nucleon scattering, whereas as $V_{ct,NNNN}^{(0)}$ acts only in $S$-waves.
For this reason the fit of low-energy constants is performed by considering individual partial waves, which goes along with a mapping of $C_{S,T}$ and $C_{1...7}^{(2)}$ to low energy constants in the spectroscopic notation.
The pertinent linear relations are given in Sec. 2.2 of Ref.~\cite{Epelbaum2005Twonucleonsystem}.	

The NLO contact potentials for the coupled channels with $\Delta$-isobars in the initial or final state are listed in \ref{sec:delta_contact_nlo}. Altogether these expressions involve 45 independent low-energy constants, which are too many to be fitted to empirical nucleon-nucleon phase shifts of e.g. the Nijmegen partial wave analysis \cite{Stoks1993Partialwaveanalysis}. 
In the actual calculation the effects of the contact terms with deltas turned out to be negligible in comparison to $V_{NNNN}^{(2)}$, since they only enter through iterations in the coupled channel equation.

The relevant contact potential $V_{ct,NNNN}^{(0)}+V_{ct,NNNN}^{(2)}$ is multiplied
with the non-local regulator function $\exp[-(p^4+p'^4)/\Lambda^4]$.

\section{Coupled channel scattering equation and nucleon-nucleon phase shifts}
\label{sec:Kadyshevsky_eq_phase_shifts}

\noindent
The chiral potentials for the coupled (NN, N$\Delta$, $\Delta$N, $\Delta\Delta$)-channels up to NLO are iterated to all orders by solving the Kadyshevsky equation \cite{Kadyshevsky1968Quasipotentialapproachand,Kadyshevsky1968Quasipotentialtypeequation} extended to four coupled channels, 

\begin{align}
K_{\nu' \nu}^{\rho' \rho,j} (p',p) =&
V_{\nu' \nu}^{\rho' \rho,j}(p',p) + \sum_{\rho'',\nu''} M_{1,\nu''} M_{2,\nu''} \fint_{0}^{\infty} \frac{\d p'' p''^2}{(2\pi)^3} \frac{V_{\nu' \nu''}^{\rho' \rho'',j}(p',p'') K_{\nu'' \nu}^{\rho'' \rho,j}(p'',p)} {E_1 E_2 (E_1 + E_2 - 2p_{0})},
\label{eq:Kadyshevsky}
\end{align}
where $j$ is the total angular momentum quantum number, $\nu$ denotes a two-baryon channel, and $\rho =(s,\ell)$ labels a partial wave.
The energy denominator is composed of the variables $p_0=\sqrt{p^2+M_N^2}$ and $E_{1,2}=\sqrt{p''^2+M_{1,2,\nu''}^2}$ and the masses of the two baryons entering the intermediate state $\nu''$ are denoted as $M_{1,\nu''}$ and $M_{2,\nu''}$.  
The Kadyshevsky equation is a modification of the non-relativistic Lippmann-Schwinger equation which includes relativistically improved kinematics. Note that the pole of the energy denominator in \cref{eq:Kadyshevsky}, which is handled by the principal value description, occurs only for the NN intermediate states.

The quantity of interest is the on-shell S-matrix for elastic nucleon-nucleon scattering. In order to obtain it, one extracts the nucleon-nucleon K-matrix $K(p)$ from the solution of the Kadyshevsky equation by setting $p'=p$ and $\nu=\nu'=NN$. The unitary S-matrix is related to the hermitian K-matrix via 
\begin{align}
S(p)= \left(1+ \iu \frac{p \mnuc^2}{16\pi^2 \sqrt{p^2 + \mnuc^2}} K(p) \right) \left(1-\iu \frac{p \mnuc^2}{16\pi^2 \sqrt{p^2 + \mnuc^2}} K(p) \right)^{-1} \; ,
\end{align}
where its matrix character refers to the partial waves.
The phase shifts and mixing angles of elastic nucleon-nucleon scattering are obtained from the S-matrix $S_{\ell \ell'}^{sj}$, where $s'=s \in \lbrace0,1\rbrace$ is the conserved total spin, via the relation for uncoupled ($\ell=\ell '=j$) partial waves
\begin{align}
S_{jj}^{sj} &= \exp(2\iu \delta_{j}^{sj})
\end{align}
and the relation for coupled ($\ell,\ell '=j\pm 1$) spin-triplet partial wave channels
\begin{align}
\begin{pmatrix}
S_{j-1j-1}^{1j} & S_{j-1j+1}^{1j} \\ S_{j+1j-1}^{1j} & S_{j+1j+1}^{1j}
\end{pmatrix}
&=
\begin{pmatrix}
\cos (2 \epsilon_{j}) \exp(2\iu \delta_{j-1}^{j}) & -\iu \sin(2 \epsilon_{j}) \exp( \iu \delta_{j-1}^{j}+ \iu \delta_{j+1}^{j}) \\
-\iu \sin(2 \epsilon_{j}) \exp( \iu \delta_{j-1}^{j}+ \iu \delta_{j+1}^{j}) & \cos (2 \epsilon_{j}) \exp(2\iu \delta_{j+1}^{j})
\end{pmatrix}
\end{align}
in the so called Stapp convention \cite{Stapp1957Phaseshiftanalysis}. The total isospin $I=0$ or $1$ is determined in each partial wave by the condition that $I+s+l$ is odd. Note that the S-matrix is unitary and symmetric, due to time-reversal invariance. The minus sign in the off-diagonal matrix elements is due to our convention in the relation between the S-matrix and the (one-pion exchange) potential.
The phase shifts $\delta_{l}^{sj}$ and mixing angles $\epsilon_{j}$ are functions of the center-of-mass momentum $p$ or the laboratory kinetic energy $T_{\text{lab}}=2p^2/\mnuc$.

\section{Results}
\label{sec:results}

In this section, we present our results for NN phase shifts and mixing angles. 
In all cases we compare results of our calculations with the purely nucleonic chiral potential and with inclusion of the coupled N$\Delta$-, $\Delta$N- and $\Delta\Delta$-channels to the Nijmegen partial wave analysis \cite{Stoks1993Partialwaveanalysis}. 

Let us first discuss our findings concerning the role of the (LO and NLO) contact interactions with deltas in the initial or final state. The evaluation of transition matrix elements shows that all NN partial waves with orbital angular momentum $\ell \leq 3$, except \pwave{3D2}, can be influenced by these contact potentials. Through iterations in the Kadyshevsky equation the $\Delta\Delta \rightarrow \Delta\Delta$ contact interaction can reach in addition some peripheral NN partial waves, e.g. the $\Delta\Delta$-wave \pwave{7S3} is coupled with the NN-waves (\pwave{3D3}, \pwave{3G3}, $\epsilon_3$), and to mention the extreme, the $\Delta\Delta$-wave \pwave{7P4} is coupled with the NN-waves (\pwave{3F4}, \pwave{3H4}, $\epsilon_4$). However, even for choices of the low-energy constants $C_{i,\Delta}^{(0)}$ and $C_{i,\Delta}^{(2)}$, which exceed the natural size by several orders of magnitude, their influence on the NN phase shifts turns out to be negligibly small. Such tiny deviations are completely covered by the variation of the cutoff in the range $\Lambda=450\dots\SI{700}{MeV}$. This features give a good reason to neglect altogether the contact interactions with deltas in the initial or final state.

Concerning the role of $1\pi$- and $2\pi$-exchange with intermediate N$\Delta$-, $\Delta$N-, $\Delta\Delta$-states, one has first the selection rule that NN partial waves with total isospin $I=0$ receive only contributions from the $\Delta\Delta$ channel. Secondly, one finds that for total isospin $I=1$ NN-waves the channels with one and two $\Delta$-isobars produce in most cases corrections of similar size.

In the following we will present and discuss our results first for the peripheral phase shifts ($\ell \geq 3$) and then for the central partial waves, where only the latter are affected by the fitted NN low-energy constants $C_{S,T}$ and $C_{1\dots 7}^{(2)}$.

\subsection{Peripheral nucleon-nucleon phase shifts}

The peripheral phase shifts are independent of the NN-contact potential, except for the \pwave{3F2}-wave and $\epsilon_2$ due to the mixing with the \pwave{3P2}-wave. However, the influence of $C^{\pwave{3P2}}$ on the $F$-wave is nearly negligible, even at the highest energy $T_{\text{lab}}\simeq \SI{300}{MeV}$. We show our calculated phase shifts and mixing angles, varying the cutoff in the range $\Lambda=450\dots\SI{700}{MeV}$.

\subsubsection{F-waves}
\begin{figure}[!htb]
	\centering
	\includegraphics[width=0.8\linewidth]{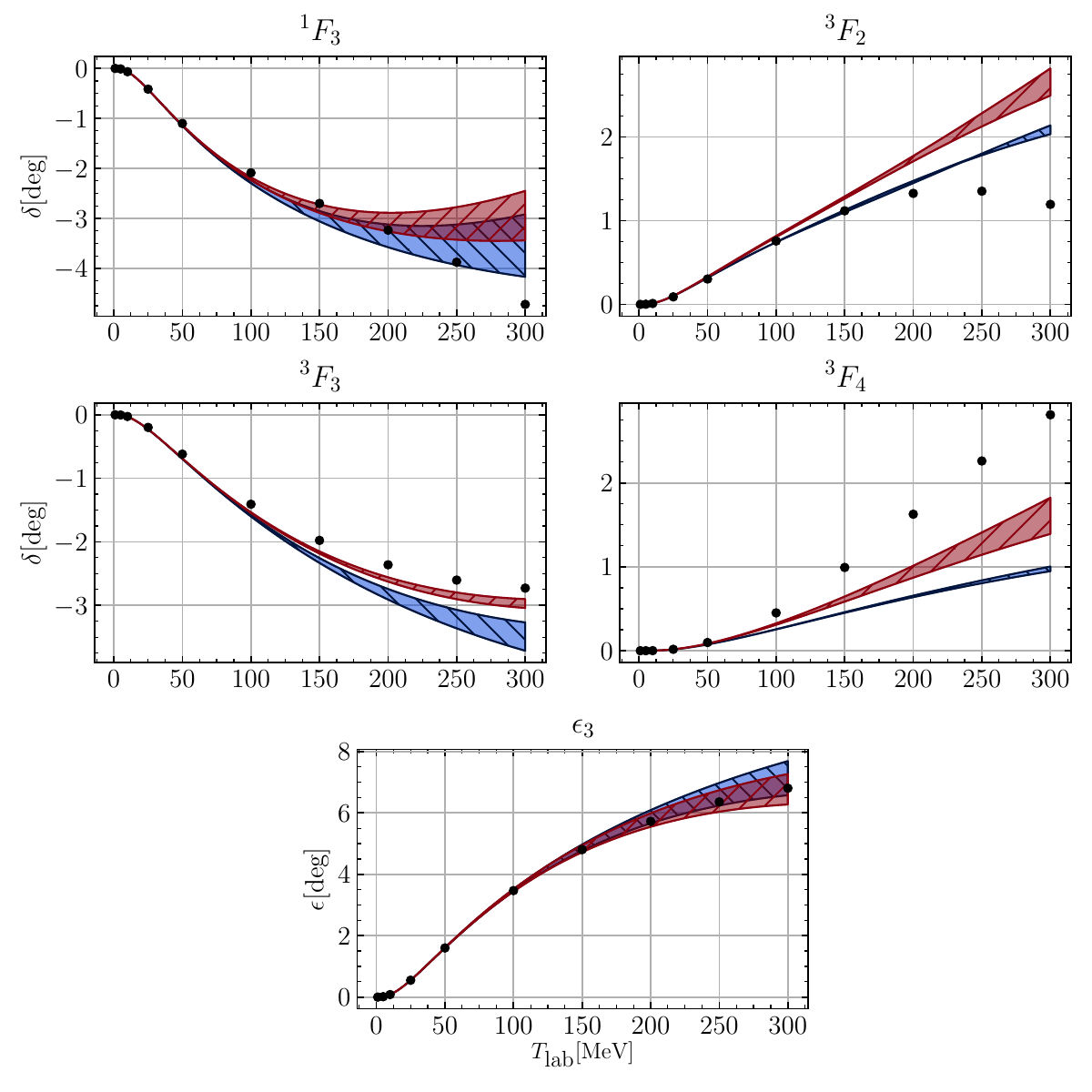}
	\caption{$F$-wave NN phase shifts and mixing angle $\epsilon_3$ versus the nucleon lab kinetic energy $T_{\text{lab}}$ for a cutoff variation $\Lambda=450\dots \SI{700}{MeV}$. The blue band (\textbackslash\textbackslash-hatching) and the red band (//-hatching) show the results of the calculation with chiral NN potentials only and with the full coupled (NN, N$\Delta$, $\Delta$N, $\Delta\Delta$) channels at NLO, respectively. The filled circles stem from the Nijmegen PWA \cite{Stoks1993Partialwaveanalysis}.}
	\label{fg:F_waves}
\end{figure}
The $F$-wave phase shifts and the mixing angle $\epsilon_3$ are shown in \cref{fg:F_waves}. With the exception of the \pwave{3F2} phase shift, the results with the coupled (N$\Delta$, $\Delta$N, $\Delta\Delta$)-channels included are closer to the Nijmegen PWA than those coming from chiral NN potentials alone. The \pwave{1F3}-wave is better described by the coupled channel approach at energies $T_{\text{lab}}< \SI{170}{MeV}$, whereas for higher energies the deviation from the empirical PWA results increases. The \pwave{3F3} phase shift improves substantially over the entire energy range $T_{\text{lab}}< \SI{300}{MeV}$, whereas for the \pwave{3F4} phase shift the corrections due to the coupled delta channels fill half of the gap between the purely nucleonic NLO calculation and the Nijmegen PWA for $T_{\text{lab}}> \SI{50}{MeV}$. The mixing angle $\epsilon_3$ can be reproduced very well. In the purely nucleonic calculation the data points lie at the lower edge of the band, whereas with coupled delta channels included the data points are located just in the middle of the band. In general the cutoff dependence for $F$-waves is rather weak. Note that by increasing the cutoff $\Lambda$, the interaction potentials get somewhat stronger and thus phase shifts tend to grow in magnitude.

\subsubsection{G-waves}
\begin{figure}[!htb]
	\centering
	\includegraphics[width=0.8\linewidth]{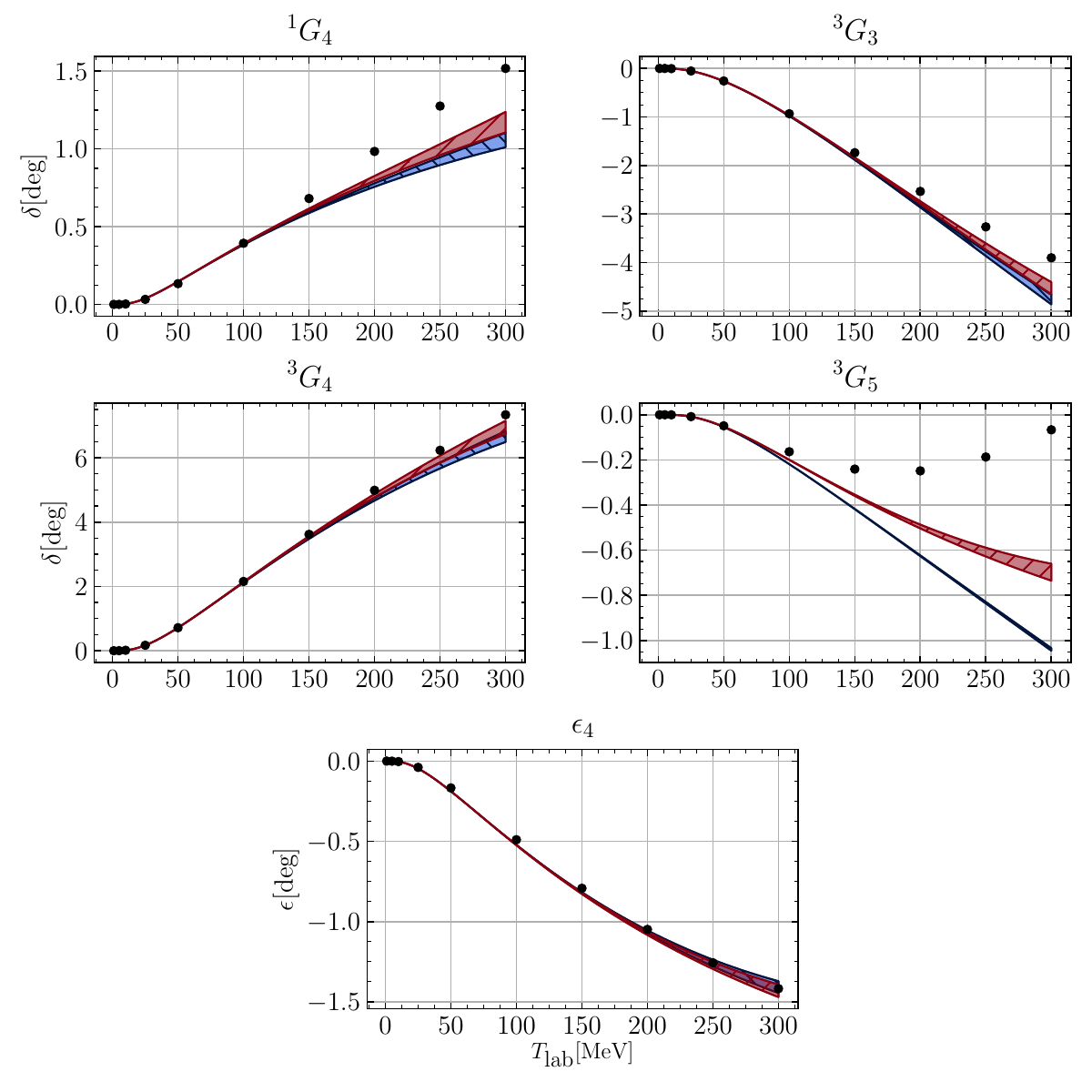}
	\caption{$G$-wave NN phase shifts and mixing angle $\epsilon_4$ versus the nucleon lab kinetic energy $T_{\text{lab}}$ for a cutoff variation $\Lambda=450\dots \SI{700}{MeV}$. For notation see \cref{fg:F_waves}.}
	\label{fg:G_waves}
\end{figure}
The $G$-wave phase shifts and the mixing angle $\epsilon_4$ are shown in \cref{fg:G_waves}. The coupled channel approach leads to some improvements in the waves \pwave{1G4}, \pwave{3G3} and \pwave{3G4} at energies $T_{\text{lab}}> \SI{200}{MeV}$. The phase shift \pwave{3G5} changes from approximately $\SI{-1.0}{\degree}$ to a phase shift of $\SI{-0.7}{\degree}$ at $\SI{300}{MeV}$ lab energy, but it still does not reproduce the curvature behavior of the Nijmegen PWA. The reproduction of this delicate feature can be achieved only at higher orders in the chiral expansion of the NN potential (see herefore the comparison of N2LO calculations with two different choices of $c_{1,3,4}$ parameters in \cref{fg:N2LO}). The cutoff dependence of the $G$-waves phase shifts has decreased significantly compared to that in lower partial waves.

\subsubsection{H-waves}
\begin{figure}[!htb]
	\centering
	\includegraphics[width=0.8\linewidth]{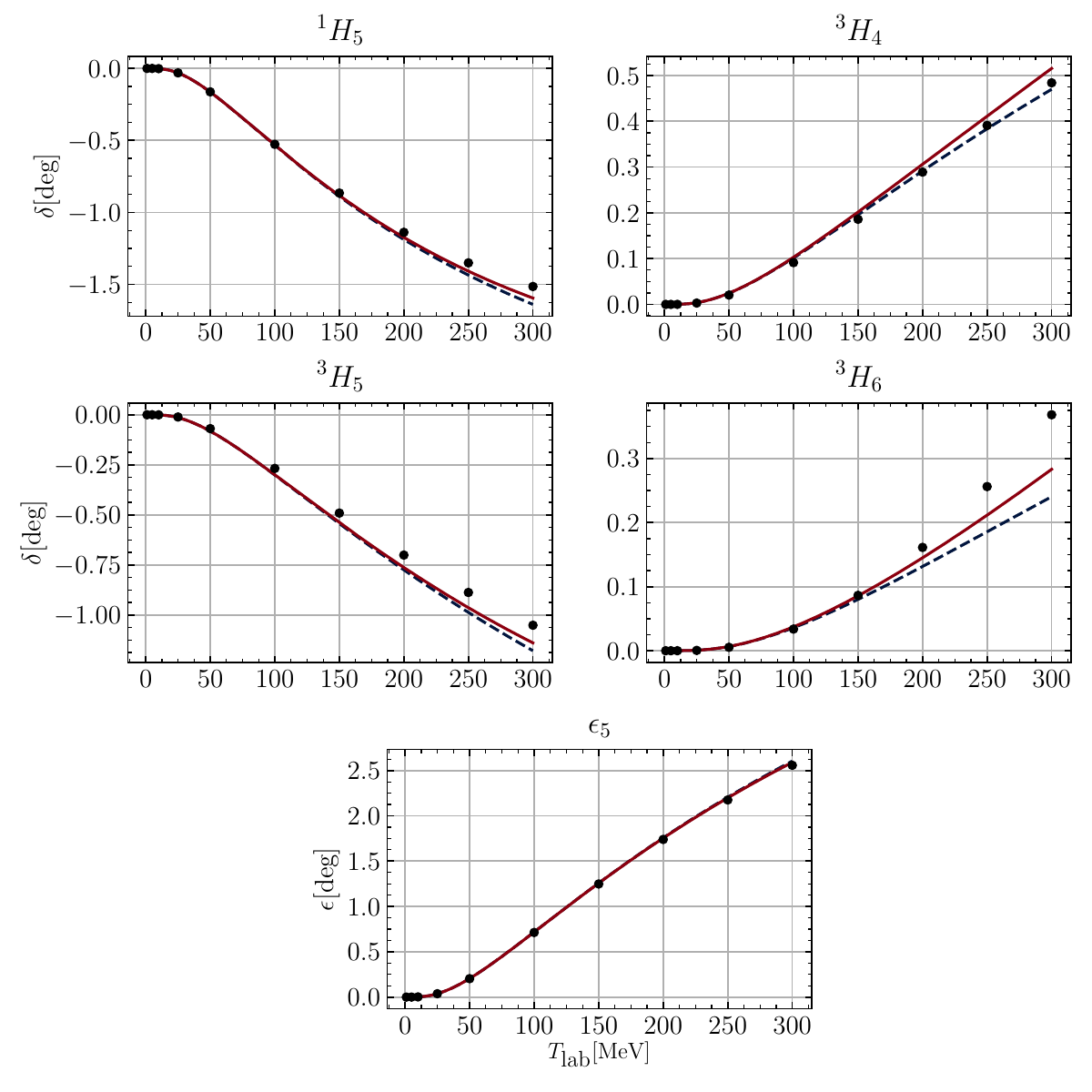}
	\caption{$H$-wave NN phase shifts and mixing angle $\epsilon_5$ versus the nucleon lab kinetic energy $T_{\text{lab}}$ for a cutoff variation $\Lambda=450\dots \SI{700}{MeV}$. The dark blue (dashed) and red (solid) line show the results of the calculation with chiral NN potentials only and with the full coupled (NN, N$\Delta$, $\Delta$N, $\Delta\Delta$) channels at NLO, respectively. The filled circles stem from the Nijmegen PWA \cite{Stoks1993Partialwaveanalysis}. There is no visible cutoff dependence in these peripheral phase shifts.}
	\label{fg:H_waves}
\end{figure}
For the $H$-waves the coupled channel approach leads to a slightly better agreement with the Nijmegen PWA at higher lab energies than the calculation of these phase shifts with purely nucleonic NLO chiral potentials, except for \pwave{3H4} as we show in \cref{fg:H_waves}. The mixing angle $\epsilon_5$ is nearly unaffected by the coupled channels. We remind that only \pwave{3H4} can have a contribution from the contact potential $V_{\Delta\Delta\Delta\Delta}^{(2)}$, but its effect is totally negligible.

\subsubsection{I-waves}
\begin{figure}[!htb]
	\centering
	\includegraphics[width=0.8\linewidth]{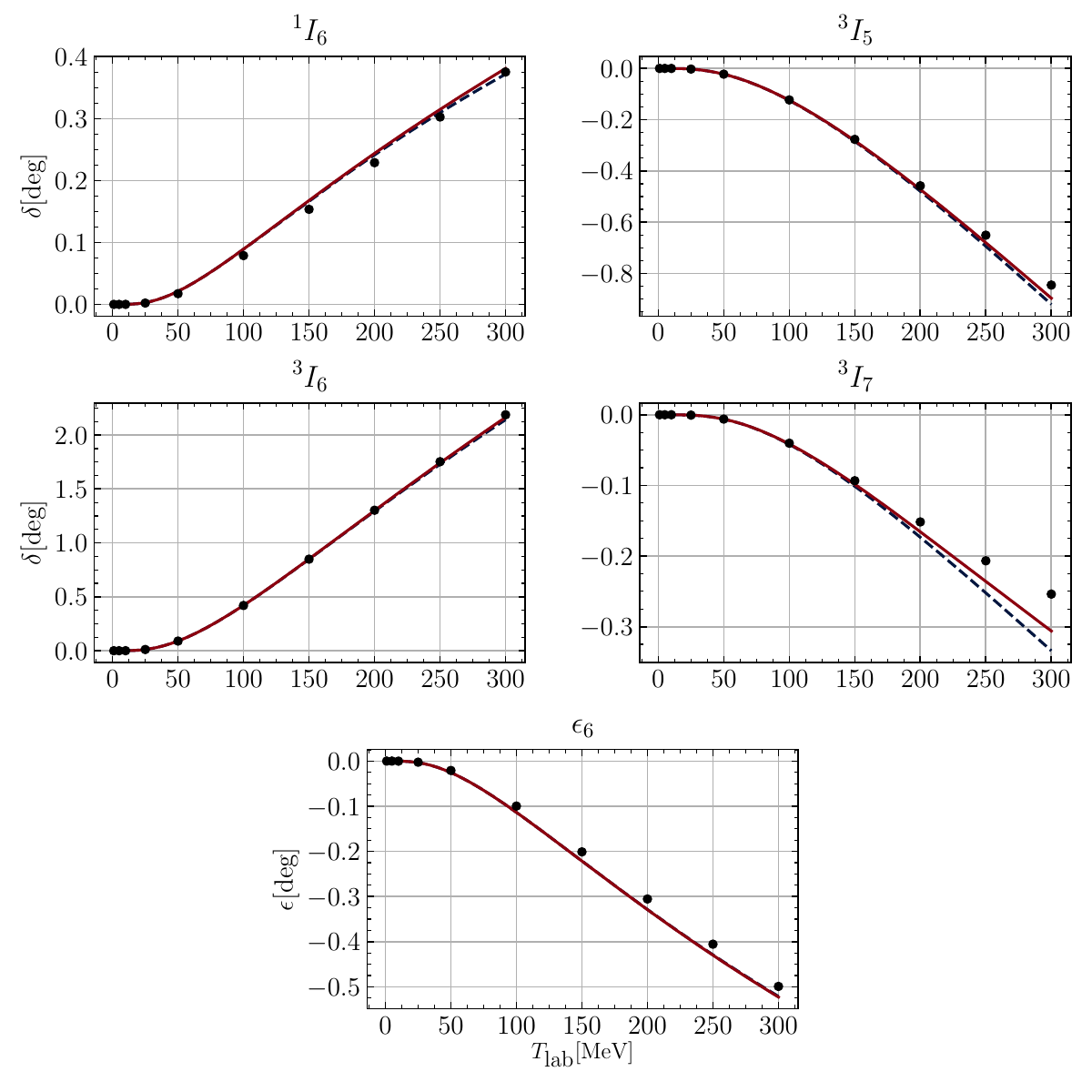}
	\caption{$I$-wave NN phase shifts and mixing angle $\epsilon_6$ versus the nucleon lab kinetic energy $T_{\text{lab}}$ for a cutoff variation $\Lambda=450\dots \SI{700}{MeV}$. For notation see \cref{fg:H_waves}.}
	\label{fg:I_waves}
\end{figure}
When taking into account the coupled (NN, N$\Delta$, $\Delta$N, $\Delta\Delta$) channels, the $I$-wave phase shifts displayed in \cref{fg:I_waves} show only very small deviations from the calculation with purely nucleonic chiral potentials. Due to the high orbital angular momentum $\ell =6$ these phase shifts are mostly dominated by the one-pion exchange between nucleons. Both the $H$-wave and $I$-wave phase shifts and the mixing angles $\epsilon_5$ and $\epsilon_6$ exhibit no visible dependence on the cutoff parameter $\Lambda$.

\subsection{Nucleon-nucleon phase shifts in low partial waves}

In the partial waves with orbital angular momentum $\ell \leq 2$, the NN contact potentials at LO and NLO affect the interaction in the $S$- and $P$-waves and also in the \pwave{3D1}-wave through the channel coupling \pwave{3S1} $\leftrightarrow$ \pwave{3D1}.
The low-energy constants $C_{S,T}$ and $C_{1\dots 7}^{(2)}$, translated into the spectroscopic notation, are determined in fits to the Nijmegen phase shifts for $T_{\text{lab}} \leq \SI{100}{MeV}$, separately for each NN partial wave channel.
In the calculation with only chiral NN potentials up to next-to-leading order we find the values of the low-energy constants $\widetilde{C}$ and $C$ listed in \cref{tb:fits_NN} for different choices of the cutoff parameter $\Lambda$. When taking into account the chiral NLO potentials for the coupled (NN, N$\Delta$, $\Delta$N, $\Delta\Delta$) channels in an analogous fit we obtain the low-energy constants $\widetilde{C}$ and $C$ collected in \cref{tb:fits_NN_coupled}.
\begin{table}[!htb]
	\centering
	\caption{Low-energy constants from fits of the purely nucleonic chiral potential to the Nijmegen PWA. The leading order constants $\widetilde{C}$ are in units of $10^4\,\si{GeV}^{-2}$, while the next-to-leading order $C$ are in units of $10^4\,\si{GeV}^{-4}$.}
	\label{tb:fits_NN}
		\begin{tabular}{lrrrrrr}
			\toprule
$\Lambda$ [MeV]			& 450 & 500 & 550 & 600 & 650 & 700 \\ \midrule
			$\widetilde{C}^{\pwave{1S0}}$ & 0.138 & 0.123 & 0.110 & 0.079 & 0.067 & 0.007 \\
			$C^{\pwave{1S0}}$     & -0.813 & -0.823 & -0.565 & -0.656 & -0.472 & -0.547 \\
			$\widetilde{C}^{\pwave{3S1}}$ & 0.173 & 0.129 & 0.170 & 0.113 & 0.048 & -1.344 \\
			$C^{\pwave{3S1}}$         & -0.769 & -0.455 & -0.801 & -0.592 & -0.599 & -1.481 \\
			$C^{\pwave{3S1}-\pwave{3D1}}$     & -0.214 & -0.295 & -0.026 & -0.178 & -0.245 & -1.004 \\
			$C^{\pwave{3P0}}$ & -0.253 & -0.309 & -0.367 & -0.441 & -0.551 & -0.750 \\
			$C^{\pwave{1P1}}$ & -0.528 & -0.445 & -0.379 & -0.325 & -0.280 & -0.242 \\
			$C^{\pwave{3P1}}$ & -0.220 & -0.190 & -0.170 & -0.157 & -0.150 & -0.147 \\
			$C^{\pwave{3P2}}$ & 0.237 & 0.206 & 0.181 & 0.160 & 0.142 & 0.126 \\ \bottomrule
		\end{tabular}
\end{table}
\begin{table}[!htb]
	\centering
	\caption{Low-energy constants from fits including coupled N$\Delta$-, $\Delta$N- and $\Delta\Delta$-channels in the chiral potential to the Nijmegen PWA. The leading order constants $\widetilde{C}$ are in units of $10^4\,\si{GeV}^{-2}$, while the next-to-leading order $C$ are in units of $10^4\,\si{GeV}^{-4}$.}
	\label{tb:fits_NN_coupled}
		\begin{tabular}{lrrrrrr}
			\toprule
$\Lambda$ [MeV]			& \multicolumn{1}{c}{450} & \multicolumn{1}{c}{500} & \multicolumn{1}{c}{550} & \multicolumn{1}{c}{600} & \multicolumn{1}{c}{650} & \multicolumn{1}{c}{700} \\ \midrule
			$\widetilde{C}^{\pwave{1S0}}$ & 0.134 & 0.119 & 0.095 & 0.046 & -0.084 & -1.234 \\
			$C^{\pwave{1S0}}$     & -0.782 & -0.773 & -0.790 & -0.855 & -1.034 & -2.064 \\
			$\widetilde{C}^{\pwave{3S1}}$ & 0.139 & 0.122 & 0.103 & 0.081 & 0.055 & 0.017 \\
			$C^{\pwave{3S1}}$  & -0.619 & -0.588 & -0.560 & -0.542 & -0.540 & -0.566 \\
			$C^{\pwave{3S1}-\pwave{3D1}}$ & -0.246 & -0.167 & -0.101 & -0.041 & 0.020 & 0.087 \\
			$C^{\pwave{3P0}}$ & -0.362 & -0.462 & -0.588 & -0.783 & -1.173 & -2.470 \\
			$C^{\pwave{1P1}}$ & -0.657 & -0.614 & -0.594 & -0.597 & -0.626 & -0.693 \\
			$C^{\pwave{3P1}}$ & -0.273 & -0.262 & -0.264 & -0.278 & -0.303 & -0.341 \\
			$C^{\pwave{3P2}}$ & 0.187 & 0.146 & 0.111 & 0.083 & 0.059 & 0.039 \\ \bottomrule
		\end{tabular}
\end{table}
One observes that for cutoffs $\Lambda\leq \SI{500}{MeV}$ both fits yield similar values for these low-energy constants. This points to the fact that the effects of the additional (N$\Delta$, $\Delta$N, $\Delta\Delta$) channels are rather small below $T_{\text{lab}}= \SI{100}{MeV}$, a feature which holds also for the peripheral partial waves.

\subsubsection{S-waves}
\begin{figure}[!htb]
	\centering
	\includegraphics[width=0.8\linewidth]{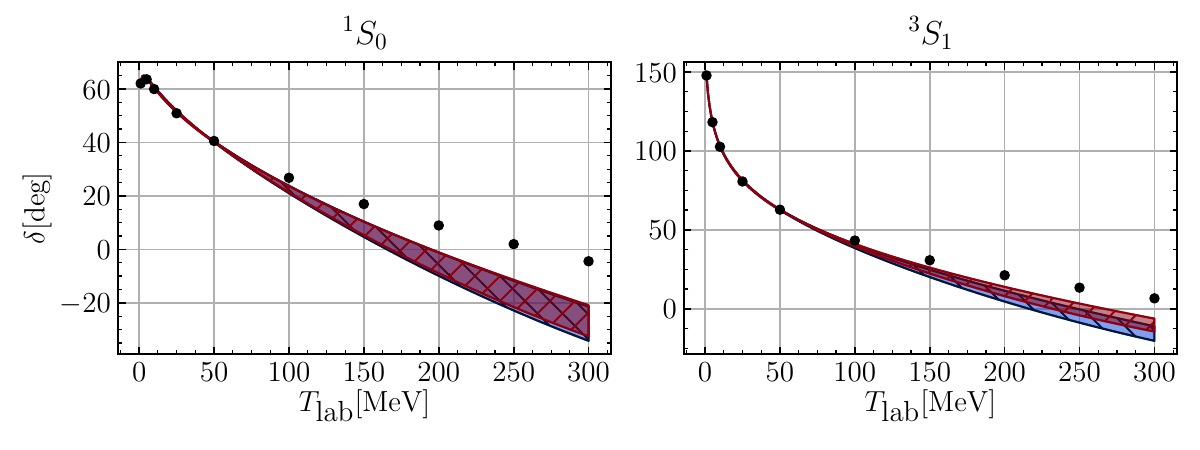}
	\caption{$S$-wave NN phase shifts versus the nucleon lab kinetic energy $T_{\text{lab}}$ for a cutoff variation $\Lambda=450\dots \SI{700}{MeV}$.  The blue band (\textbackslash\textbackslash-hatching) and the red band (//-hatching) show the results of the calculation with chiral NN potentials only and with the full coupled (NN, N$\Delta$, $\Delta$N, $\Delta\Delta$) channels at NLO, respectively. The filled circles stem from the Nijmegen PWA \cite{Stoks1993Partialwaveanalysis}.}
	\label{fg:S_waves}
\end{figure}
We have seen that with increasing orbital angular momentum $\ell$, the cutoff dependence of the peripheral phase shifts decreases rapidly. However, in the low partial waves the cutoff plays a major role for the two-pion exchange potentials but this regularization dependence is balanced in the $S$- and $P$-waves to a large extent by the NN contact potentials with their adjustable parameters.
The calculated results for the $S$-wave phase shifts are shown in \cref{fg:S_waves}. 
Of course, one has to perform separate fits in the coupled (NN, N$\Delta$, $\Delta$N, $\Delta\Delta$)-channel approach and the purely nucleonic calculation in order to account for the differences of the $1\pi$- and $2\pi$-exchange potentials. Due to the strong influence of the low-energy constants $\widetilde{C}^{\pwave{1S0}}$, $C^{\pwave{1S0}}$, $\widetilde{C}^{\pwave{3S1}}$, $C^{\pwave{3S1}}$ and $C^{\pwave{3S1}-\pwave{3D1}}$ the effect of the coupled (N$\Delta$, $\Delta$N, $\Delta\Delta$)-channels is almost negligible, as one can see from the overlapping bands in \cref{fg:S_waves}. The fitted values of these five low-energy constants can be found in \cref{tb:fits_NN,tb:fits_NN_coupled} for the considered cutoff range $\Lambda=450\dots \SI{700}{MeV}$. The better reproduction of the \pwave{3S1} phase shift compared to the \pwave{1S0} phase shift is a typical feature of NLO and N2LO calculations. The convergence of the chiral expansion for the nucleon-nucleon $S$-waves is nicely illustrated in Fig.~3 of Ref.~\cite{Epelbaum2005Twonucleonsystem}.

\subsubsection{P-waves}
\begin{figure}[!htb]
	\centering
	\includegraphics[width=0.8\linewidth]{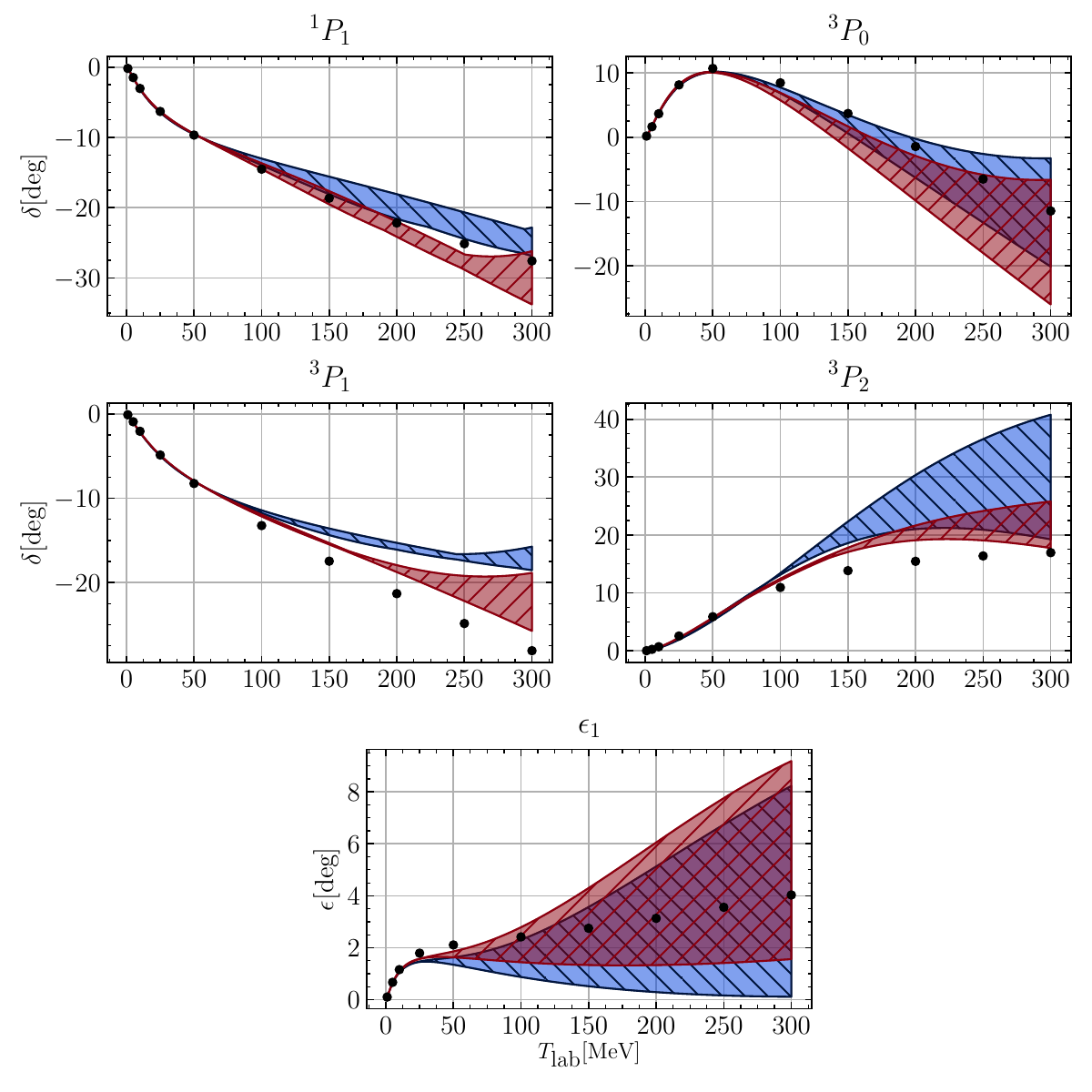}
	\caption{$P$-wave NN phase shifts and mixing angle $\epsilon_1$ versus the nucleon lab kinetic energy $T_{\text{lab}}$ for a cutoff variation $\Lambda=450\dots \SI{700}{MeV}$. For notation see \cref{fg:S_waves}.}
	\label{fg:P_waves}
\end{figure}
The calculated results for the $P$-wave phase shifts and the mixing angle $\epsilon_1$ are shown in \cref{fg:P_waves}. These quantities are influenced by the low-energy constants $C^{\pwave{1P1}}$, $C^{\pwave{3P0}}$, $C^{\pwave{3P1}}$, $C^{\pwave{3P2}}$ and $C^{\pwave{3S1}-\pwave{3D1}}$. The corresponding fitted values are listed in \cref{tb:fits_NN,tb:fits_NN_coupled}. 
The \pwave{3P0} phase shift obtained in the coupled (NN, N$\Delta$, $\Delta$N, $\Delta\Delta$) channel approach deviates a bit more from the Nijmegen data points than that of the purely nucleonic calculation for $T_{\text{lab}} > \SI{150}{MeV}$.
For the \pwave{1P1} phase shift the inclusion of the coupled (N$\Delta$, $\Delta$N, $\Delta\Delta$) channels leads to a better agreement with the Nijmegen PWA up to $T_{\text{lab}} \approx \SI{200}{MeV}$. The \pwave{3P1} and \pwave{3P2} phase shifts receive corrections which move the rather narrow bands towards the data and the cutoff dependence is reduced significantly for the \pwave{3P2} phase shift. 
At low lab kinetic energies $T_{\text{lab}}<\SI{50}{MeV}$ the mixing angle $\epsilon_1$ comes out closer to the empirical values in the coupled channel approach, but with increasing energies the cutoff dependence of $\epsilon_1$ grows in the same way for the calculation with and without coupled channels, such that the two bands overlap.

\subsection{D-waves}
\begin{figure}[!htb]
	\centering
	\includegraphics[width=0.8\linewidth]{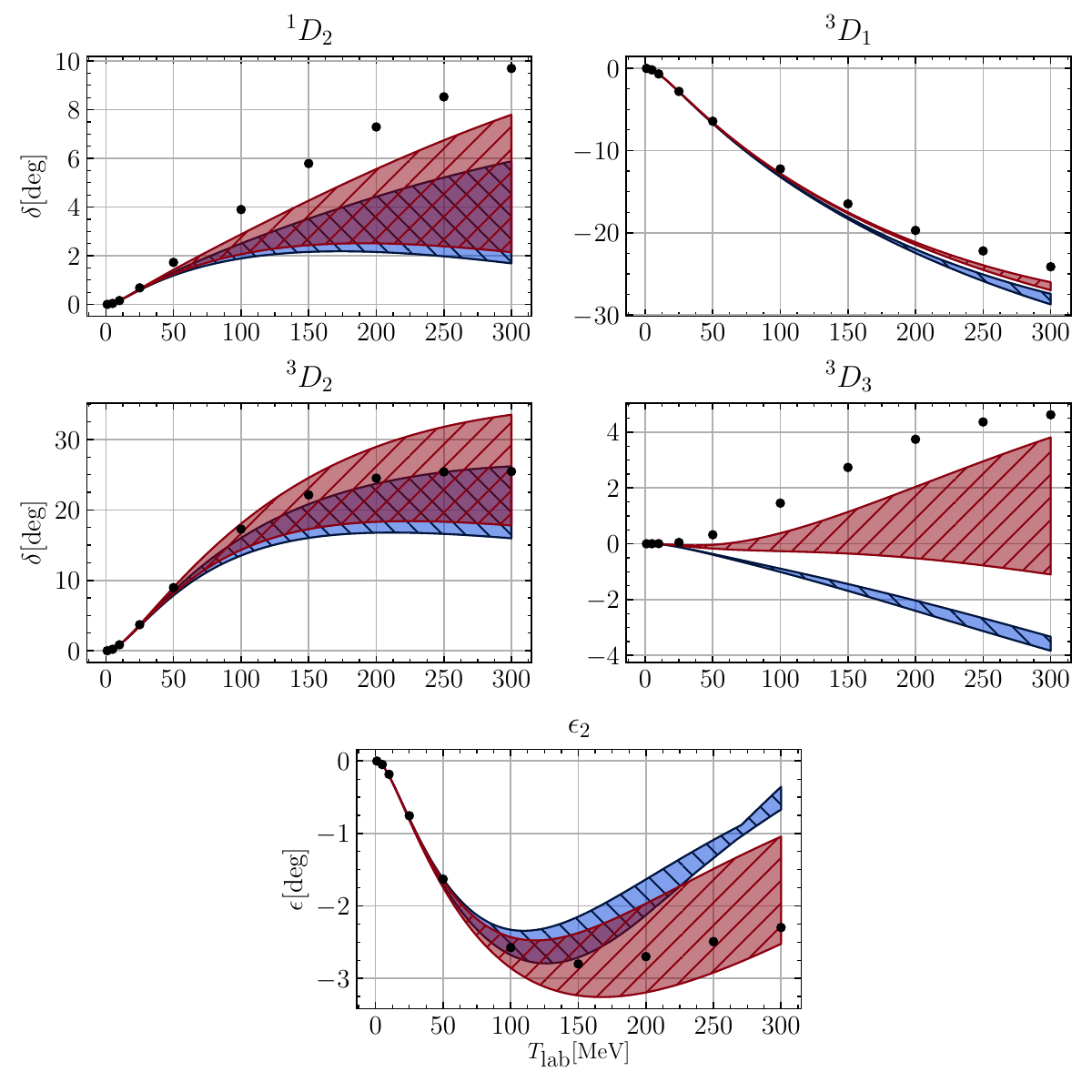}
	\caption{$D$-wave NN phase shifts and mixing angle $\epsilon_2$ versus the nucleon lab kinetic energy $T_{\text{lab}}$ for a cutoff variation $\Lambda=450\dots \SI{700}{MeV}$. For notation see \cref{fg:S_waves}.}
	\label{fg:D_waves}
\end{figure}
The calculated results for the $D$-wave phase shifts and the mixing angle $\epsilon_2$ are shown in \cref{fg:D_waves}. Except for the \pwave{3D1} phase shift, there is no influence of the NN contact potentials on the $D$-waves. In all cases the corrections generated by the coupled (NN, N$\Delta$, $\Delta$N, $\Delta\Delta$) channels tend into the right direction. 
For the \pwave{3D2} phase shift and the mixing angle $\epsilon_2$, the Nijmegen PWA results lie within the band obtained in the coupled (NN, N$\Delta$, $\Delta$N, $\Delta\Delta$) channel approach. The bands resulting from the cutoff variation $\Lambda=450\dots \SI{700}{MeV}$ widen in most cases with increasing $T_{\text{lab}}$. Only for the \pwave{3D1} phase shift, which is influenced by the part of the NN contact potential with low-energy constants $\widetilde{C}^{\pwave{3S1}}$, $C^{\pwave{3S1}}$ and $C^{\pwave{3S1}-\pwave{3D1}}$ through the channel coupling \pwave{3S1} $\leftrightarrow$ \pwave{3D1}, this cutoff dependence is strongly counterbalanced. The result of the coupled (NN, N$\Delta$, $\Delta$N, $\Delta\Delta$) channel approach for the \pwave{3D3} phase shift represents a significant improvement over that of the calculation with the purely nucleonic chiral NN potential at NLO. Especially, for higher cutoffs $\Lambda > \SI{650}{MeV}$ the calculated phase shifts lie close to the empirical points from the Nijmegen PWA.

\subsection{Deuteron properties}

In this subsection, we briefly touch upon the deuteron properties. When including the $\Delta\Delta$-channel with total isospin $I=0$, the bound state equation for the deuteron reads (in our sign-convention for the momentum-space potential $V$):
\begin{align}	
	\Phi_{\nu}^{\rho}(p) = \frac{1}{B_d + 2 \Delta \xi_{\nu}  +  p^2/M_{\nu}} \sum_{\nu' \rho'} \int_{0}^{\infty} \frac{\d p' p'^2}{(2\pi)^3} V_{\nu\nu'}^{\rho\rho',1}(p,p') \Phi_{\nu'}^{\rho'}(p')\,,
\end{align}
with $B_d$ the (positive) deuteron binding energy and coefficients $\xi_{NN}=0$, $\xi_{\Delta\Delta}=1$, that take care of twice the $\Delta$N mass splitting in the $\Delta\Delta$-channel. In this extended scheme, the deuteron wave function $\Phi_{\nu}^{\rho}(p)$ has six components, namely the components (\pwave{3S1}, \pwave{3D1}) for the NN-state and the components (\pwave{3S1}, \pwave{3D1}, \pwave{7D1}, \pwave{7G1}) for the $\Delta\Delta$-state.

\begin{figure}[!htb]
	\centering
	\includegraphics[width=0.5\linewidth]{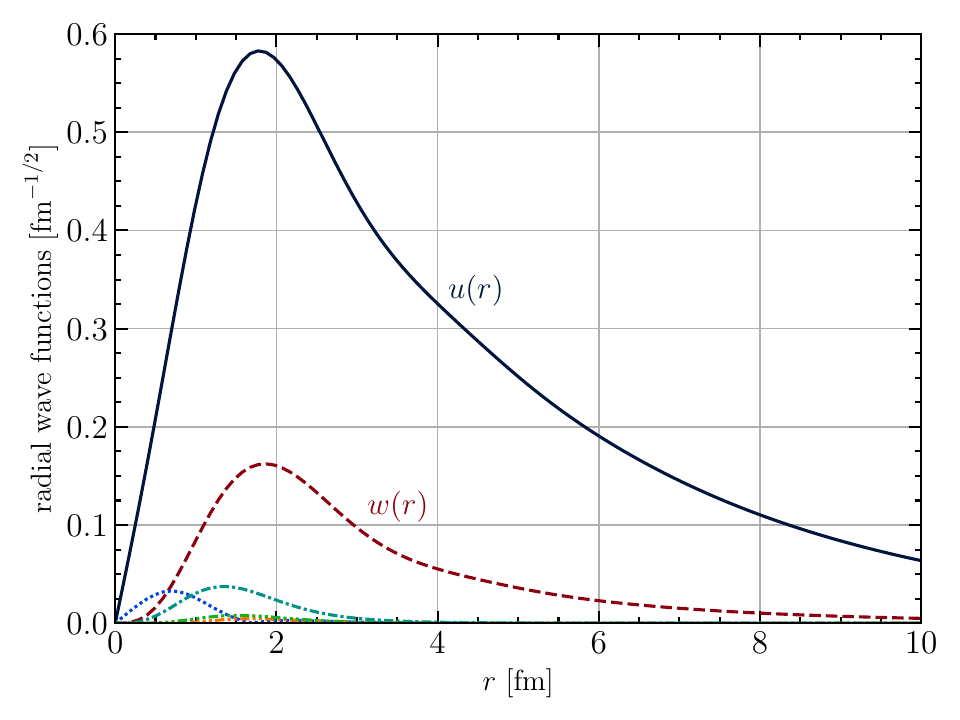}
	\caption{Radial deuteron wave functions of the NN $S$-wave $u(r)$ and NN $D$-wave $w(r)$, together with the four small components for the $\Delta\Delta$-state for a cutoff $\Lambda=\SI{450}{MeV}$.}
	\label{fg:deuteron}
\end{figure}

Let us consider the situation with the lowest cutoff $\Lambda=\SI{450}{MeV}$ and all low-energy constants fixed, see second column in \cref{tb:fits_NN_coupled}. Without any fitting, one obtains a deuteron binding energy of $B_d=\SI{3.11}{MeV}$, which is certainly too large. The results for other deuteron properties, such as the $D$-state probability $P_D$, the quadrupole moment $Q_d$, the matter radius $r_d$, the asymptotic $S$-state normalization $A_S$, and the asymptotic $D$/$S$-ratio $\eta$ are given in \cref{tb:deuteron}. 
\begin{table}[!htb]
	\caption{Deuteron properties as explained in the text. The collection of empirical values is taken from Ref.~\cite{Reinert2017Semilocalmomentumspace}.}
	\label{tb:deuteron}
	\centering
	\begin{tabular}{ccccccc}
		\toprule
		& $B_d\; [\si{MeV}]$ & $P_D\; [\si{\percent}]$ & $Q_d\; [\si{fm^2}]$ & $r_d\; [\si{fm}]$ & $A_S\; [\si{fm^{-1/2}}]$ & $\eta$ \\ \midrule
		including $\Delta\Delta$ & 3.11 & 4.97 & 0.257 & 1.71 & 0.990 & 0.0305 \\ 
		purely NN & 2.66 & 4.09 & 0.295 & 1.80 & 0.909 & 0.0281 \\ 
		empirical & 2.22 & 3.6$-$5.3& 0.286 & 1.98 & 0.885 & 0.0256 \\ \bottomrule
	\end{tabular} 
\end{table}
The corresponding radial wave functions, obtained by transformation of the six components of $\Phi_{\nu}^{\rho}(p)$ with spherical Bessel functions $j_{0,2,4}(pr)$, are shown in \cref{fg:deuteron} for distances up to $r=\SI{10}{fm}$. One observes that the $\Delta\Delta$ components are very small, with a contribution of only $0.24\%$ to the normalization. For comparison we report also on the situation with pure NN potentials and low-energy constants as in the second column in \cref{tb:fits_NN}. In this case the deuteron binding energy comes out as $B_d=\SI{2.66}{MeV}$, and the results for the other deuteron properties are given in \cref{tb:deuteron} together with empirical values. One makes the observation that in NLO calculations (with and without coupled N$\Delta$-channels) especially the result for the mixing angle $\epsilon_1$ deteriorates, when taking into account the deuteron binding energy $B_d=2.22$\,MeV into the fit of the low-energy constants.

\section{Next-to-next-to-leading order effects}
\label{sec:N2LO}
\begin{figure}[!htb]
	\centering
	\includegraphics[width=\textwidth]{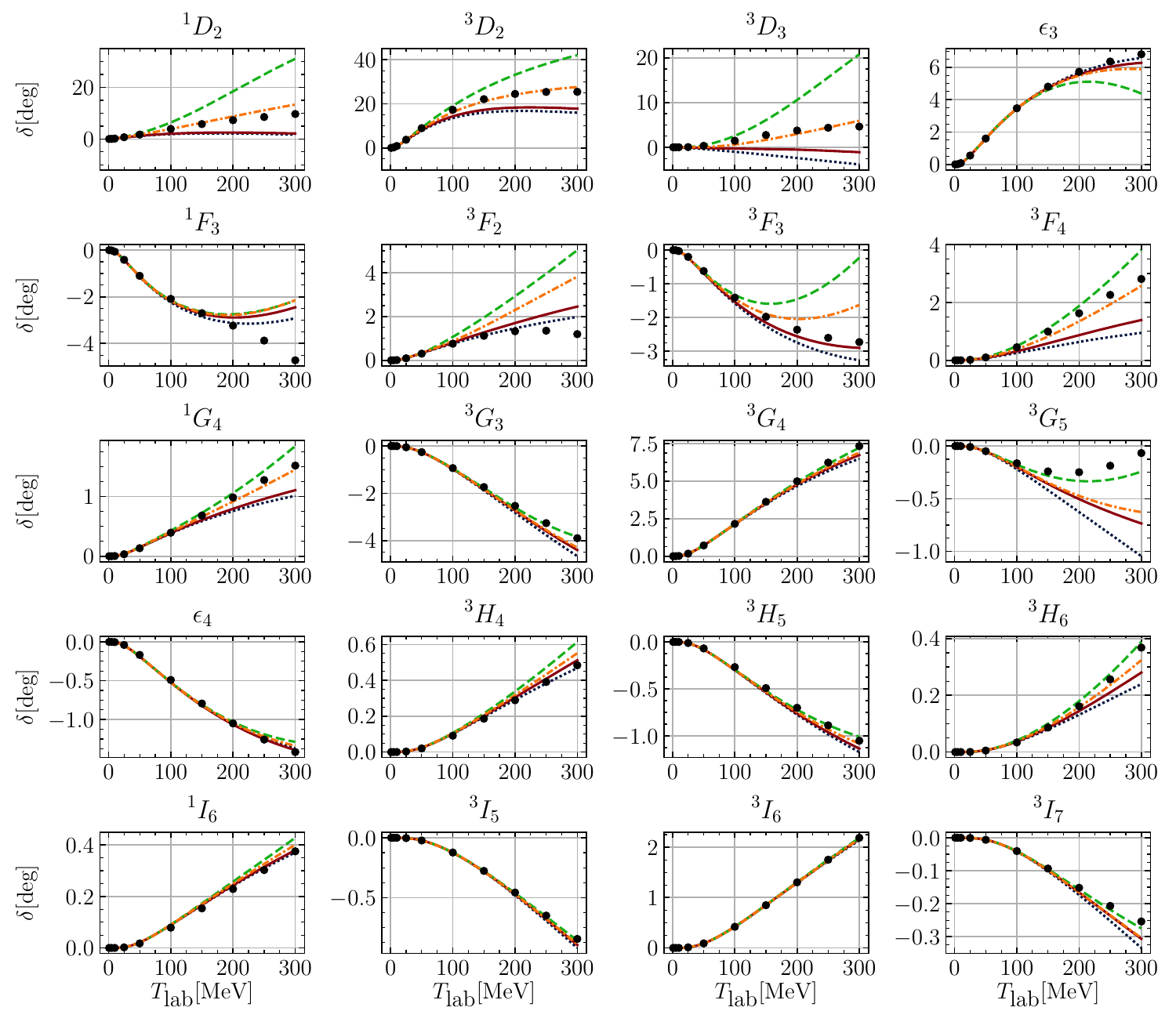}
	\caption{Selected NN phase shifts and mixing angles for a cutoff $\Lambda=\SI{450}{MeV}$. Red solid curves: NLO calculation with coupled N$\Delta$-channels, dark blue dotted curves: NLO calculation with chiral NN potentials only, green dashed curves: N2LO calculation with empirical low-energy constants $c_{1,3,4}$, orange dash-dotted curves: N2LO calculation with $c_1=0$, $c_3 = -2c_4 = - \ga^2/(2\Delta)$.}
	\label{fg:N2LO}
\end{figure}
In chiral effective field theory with pions and nucleons as active degrees of freedom only, the effects of the $\Delta(1232)$-resonance are hidden in the low-energy constants $c_{2,3,4}$ of higher derivative $2\pi$NN-vertices \cite{Kaiser1997Peripheralnucleonnucleon}. It is therefore interesting to compare the $\Delta$-less theory for the chiral NN potential at N2LO, which includes additional $2\pi$-exchange contributions proportional to $c_{1,2,3,4}$, to our coupled (NN, N$\Delta$, $\Delta$N, $\Delta\Delta$) channel approach.

At next-to-next-to-leading order we take the spectral functions of the isoscalar central potential $V_C(q)$ and isovector spin-spin and tensor potential $[W_S(q)\, \vec\sigma_1 \cdot \vec\sigma_2 + W_T(q)\, \vec\sigma_1 \cdot \vec q \, \, \vec\sigma_2 \cdot \vec q \,] \vec\tau_1 \cdot \vec\tau_2 $ from Ref.~\cite{Kaiser1997Peripheralnucleonnucleon}, which read
\begin{align} 
\Im V_C &=\frac{3g_A^2 }{ 64 f_\pi^4} \left[2m_\pi^2( 2c_1-c_3)+\mu^2  c_3 \right]\frac{2 \mpi^2 - \mu^2}{\mu}  \,, \nonumber\\
\Im W_T &=\frac{1}{ \mu^2} \Im W_{S} 
=\frac{g_A^2 }{ 128 f_\pi^4}\, c_4  \, \frac{4 \mpi^2 - \mu^2}{\mu}\,.
\label{eq:N2LO}
\end{align} 
The potentials $V_C(q)$ and $W_{S,T}(q)$ are again constructed through the regularized spectral representation in \cref{eq:regulator_2pi}.
We employ the (central) values of the low-energy constants $c_{1,3,4}$ as obtained in Ref. \cite{Buettiker2000Pionnucleonscattering} from a fit to pion-nucleon scattering inside the Mandelstam triangle: $c_1 = -0.81\si{GeV}^{-1}$, $c_3 = -4.69 \si{GeV}^{-1}$ and $c_4 = 3.40 \si{GeV}^{-1}$, which were also used for elastic nucleon-nucleon scattering in Ref. \cite{Epelbaum2015Improvedchiralnucleon}. In addition, we consider also a second set of values: $c_1=0$, $c_3 = -2c_4 = - \ga^2/(2\Delta) = \SI{-2.84}{GeV}^{-1}$, that represent the parts arising solely from the $\Delta(1232)$-resonance excitation \cite{Bernard1997Aspectschiralpion,Krebs2007NuclearforcesDelta}. The pion-nucleon LEC $c_1$ is related to explicit chiral symmetry breaking and receives no contribution from the $\Delta$-resonance. 

The results for selected NN phase shifts and mixing angles are shown in \cref{fg:N2LO} for a fixed cutoff $\Lambda=\SI{450}{MeV}$. The dashed curves correspond to the calculation with empirical values of $c_{1,3,4}$ and the dash-dotted curves refer to the second set. One observes that the results of our coupled (NN, N$\Delta$, $\Delta$N, $\Delta\Delta$) channel approach lie in between the NLO and N2LO calculations with purely nucleonic potentials for both choices of the low-energy constants $c_{1,3,4}$. As expected, the second set of LECs gives smaller changes of the phase shifts compared to NLO than the first set. 

For the phase shifts in the partial waves \pwave{1F3}, \pwave{3G3}, \pwave{3G4}, \pwave{3G5}, \pwave{3H6}, and in all $I$-waves, as well as for the mixing angles $\epsilon_3$ and $\epsilon_4$, the result of the coupled (NN, N$\Delta$, $\Delta$N, $\Delta\Delta$) channel approach agrees reasonably well with the N2LO calculation for set two. For the $D$-wave phase shifts the coupled (NN, N$\Delta$, $\Delta$N, $\Delta\Delta$) channel approach produces weaker effects than the N2LO calculation for set two. On the other hand, the pion-nucleon low-energy constants $c_{1,3,4}$ (set two) yield too strong effects in the \pwave{3F2}-, \pwave{3F3}- and \pwave{3H4}-waves, whereas the results of the coupled channel approach agree much better with the empirical phase shifts.
The coupled (NN, N$\Delta$, $\Delta$N, $\Delta\Delta$) channel dynamics generates (through infinite iterations) higher order corrections to the interaction strength represented by the low-energy constants  $c_{3,4}$, and apparently these corrections work in the opposite direction (in a specific way for each partial wave). Such a reduction of the delta dynamics encoded in $c_{3,4}$ is favorable for some partial waves but disfavorable for others. This mixed findings point to the need for N3LO or even N4LO calculations \cite{Epelbaum2015Improvedchiralnucleon,Epelbaum2015Precisionnucleonnucleon,Entem2015Peripheralnucleonnucleon} in order to get an accurate description of elastic nucleon-nucleon scattering in all partial waves.

\section{Summary and conclusions}
\label{sec:conclusion} 
In this work, we have calculated elastic nucleon-nucleon scattering in chiral effective field theory taking into account the coupled (NN, N$\Delta$, $\Delta$N, $\Delta\Delta$)-channels. The interaction potentials arising from $1\pi$- and $2\pi$-exchange as well as contact terms have been derived up to next-to leading order. This extension to coupled channels required the evaluation of six additional one-pion exchange diagrams at leading order and about 60 two-pion exchange diagrams at next-to-leading order. The pertinent two-baryon contact interactions for this approach have been classified at leading order and at next-to-leading order, where they provide contact potentials that either are momentum independent or depend quadratically on momenta.

Analytic expressions have been derived for the spectral functions following from the two-pion exchange diagrams, considering the seven independent combinations of initial and final (NN, N$\Delta$, $\Delta$N, $\Delta\Delta$)-states and the nine possible intermediate states with two, one, or zero baryons corresponding to box, triangle, or bubble diagrams. The reducible components of the planar box diagrams have been identified and excluded from the potential. The crucial rule was: The irreducible parts of the N$\Delta$ and the $\Delta\Delta$ planar box diagram are \textit{equal} and coincide with the \textit{negative} of the N$\Delta$ crossed box diagram. 
The $2\pi$-exchange potentials depending on the momentum transfer $q$ have been obtained from the analytically calculated spectral functions through regularized dispersion integrals employing the local regulator of Ref.~\cite{Reinert2017Semilocalmomentumspace}.

Based on these NLO chiral (transition) potentials for total isospin $I=0,1$ and the Kadyshevsky equation the phase shifts and mixing angles of elastic nucleon-nucleon scattering have been calculated. Since the contact potentials with deltas in the initial or final state had a totally negligible influence on the NN phase shifts, compared to the effect of the cutoff variation $\Lambda=400\dots \SI{700}{MeV}$, we have dropped these contact interactions and fitted only the nine low-energy constants $C_{S,T}$ and $C_{1\dots 7}$, which parameterize the NN contact potential up to NLO.
In comparison to the calculation with purely nucleonic chiral potentials at next-to-leading order, the coupled (NN, N$\Delta$, $\Delta$N, $\Delta\Delta$) channel approach leads in most partial waves to an improved description of the empirical phase shifts and mixing angles from the Nijmegen PWA. However, the corrections arising from the coupled (N$\Delta$, $\Delta$N, $\Delta\Delta$) channels are still too weak at higher lab kinetic energies $T_{\text{lab}}$, especially for \pwave{1D2}, \pwave{3D3}, \pwave{3F4}, \pwave{3G5} and \pwave{3H6}. The inclusion of the coupled channels has also led to a reduction of the regularization cutoff dependence in the partial waves \pwave{1P1}, \pwave{3P2}, \pwave{3D1}, \pwave{1F3}, and \pwave{3F3}, whereas it remained constant or increased in other partial waves with $\ell \leq 4$. 

We have also performed a N2LO calculation where the effects from $\Delta(1232)$-excitations are encoded in the pion-nucleon low-energy constants $c_{1,3,4}$ and the corresponding $2\pi$-exchange NN potentials.
We have found that in most peripheral partial waves with $\ell \geq 4$, the N2LO contribution attributed to the $\Delta$-isobar can be obtained in the same way with coupled (NN, N$\Delta$, $\Delta$N, $\Delta\Delta$) channels at NLO. For $D$- and $F$-waves the situation was mixed. If the effects of $c_{1,3,4}$ were too strong, the coupled channel dynamics generated higher order corrections which improved the description. However, there were also partial waves where such a reduction was unfavorable. These ambivalent findings point to the need for N3LO or even N4LO calculations \cite{Epelbaum2015Improvedchiralnucleon,Epelbaum2015Precisionnucleonnucleon,Entem2015Peripheralnucleonnucleon} in order to get an accurate description of elastic nucleon-nucleon scattering in all partial waves.

\section*{Acknowledgments}
We thank Patrick Reinert for providing a code for the numerical solution of the Kadyshevsky equation and the fits of the low-energy constants. 
We are also grateful to Wolfram Weise for informative discussions.

\appendix

\section{Spin and isospin matrices and relations}
\label{sec:spin_and_isospin_matrices}
In this appendix the explicit forms and properties of the spin (transition) matrices $\sigma^i$, $S^i$, $S^{i\dagger}$ and $\Sigma^i$ are collected. All definitions and relations apply in exactly the same way to the isospin (transition) matrices $\tau^i$, $T^i$, $T^{i\dagger}$ and $\Theta^i$.
\begin{itemize}
	\item The spin-$\frac{1}{2}$ matrices are the usual Pauli matrices $\sigma^{i}$
	\begin{align}
	\sigma^{1}=
	\begin{pmatrix}
	0 & 1 \\
	1 & 0
	\end{pmatrix}
	,~
	\sigma^{2}=
	\begin{pmatrix}
	0 & -\iu \\
	\iu & 0
	\end{pmatrix}
	,~
	\sigma^{3}=
	\begin{pmatrix}
	1 & 0 \\
	0 & - 1
	\end{pmatrix}.
	\label{eq:sigma_matrices}
	\end{align}
	\item The transition matrices for spin-$\frac{3}{2}$ to spin-$\frac{1}{2}$ read
	\begin{gather}
	S^{1}= \frac{1}{\sqrt{6}}
	\begin{pmatrix}
	-\sqrt{3} & 0 & 1 & 0 \\
	0 & - 1 & 0 & \sqrt{3} 
	\end{pmatrix}
	,~
	S^{2}= \frac{-\iu}{\sqrt{6}}
	\begin{pmatrix}
	\sqrt{3} & 0 &  1 & 0 \\
	0 &  1 & 0 & \sqrt{3} 
	\end{pmatrix}
	, ~
	S^{3}= \frac{1}{\sqrt{6}}
	\begin{pmatrix}
	0 & 2 & 0 & 0 \\
	0 & 0 & 2 & 0 
	\end{pmatrix},
	\label{eq:Spin_transition_matrices}
	\end{gather}
	and their hermitian conjugates $S^{i \dagger}$ serve for the reverse transition spin-$\frac{1}{2}$ to spin-$\frac{3}{2}$.
	\item The spin-$\frac{3}{2}$ matrices take the form
	\begin{gather}
	\Sigma^{1}=
	\begin{pmatrix}
	0 & \sqrt{3} & 0 & 0 \\
	\sqrt{3} & 0 & 2 & 0 \\
	0 & 2 & 0 & \sqrt{3} \\
	0 & 0 & \sqrt{3} & 0
	\end{pmatrix}
	,~
	\Sigma^{2}=
	\begin{pmatrix}
	0 & -\sqrt{3}\iu & 0 & 0 \\
	\sqrt{3}\iu & 0 & - 2 \iu& 0 \\
	0 & 2\iu & 0 & -\sqrt{3} \iu \\
	0 & 0 & \sqrt{3} \iu & 0
	\end{pmatrix}
	,~
	\Sigma^{3}=
	\begin{pmatrix}
	3 & 0 & 0 & 0 \\
	0 & 1 & 0 & 0 \\
	0 & 0 &- 1 & 0 \\
	0 & 0 & 0 &- 3
	\end{pmatrix}.
	\label{eq:Sigma_matrices}
	\end{gather}
	\item We use additional spin-matrices defined in Ref.~\cite{Haidenbauer2017Scatteringdecupletbaryons}, which are symmetric in the multiple indices $i,j,k \in \lbrace 1,2,3 \rbrace$,
	\begin{align}
	S^{ij \dagger} =& - \frac{1}{\sqrt{6}} \bigl( S^{i \dagger} \sigma^{j} + S^{j \dagger} \sigma^{i} \bigr) \; ,  \nonumber \\
	S^{ij}  =&  - \frac{1}{\sqrt{6}} \bigl( \sigma^{i} S^{j} + \sigma^{j} S^{i}  \bigr) \; , \nonumber \\
	\Sigma^{ij}  =& \frac{1}{8} \bigl( \Sigma^{i} \Sigma^{j} + \Sigma^{j} \Sigma^{i} - 10 \delta^{ij} \1 \bigr) = \delta^{ij} \1 - \frac{3}{2} \bigl( S^{i \dagger} S^{j} + S^{j \dagger} S^{i} \bigr) \; , \nonumber \\
	\Sigma^{ijk}  =&	\frac{1}{36 \sqrt{3}} \biggl( 5 (\Sigma^i \Sigma^j \Sigma^k + \Sigma^k \Sigma^i \Sigma^j + \Sigma^j \Sigma^k \Sigma^i + \Sigma^i \Sigma^k \Sigma^j \nonumber\nonumber \\
	&+ \Sigma^j \Sigma^i \Sigma^k + \Sigma^k \Sigma^j \Sigma^i ) - 82 ( \Sigma^i \delta^{jk}  + \Sigma^j \delta^{ik} + \Sigma^k \delta^{ij} )\biggr)  \; .
	\label{eq:spin_matrices_tensors}
	\end{align}
	\item The products of two matrices fulfill the following relations
	\begin{align}
	\sigma^{i}\sigma^{j}&= \delta^{ij} \1+ \iu \epsilon^{ijk} \sigma^{k}  \; , \nonumber \\
	S^{i \dagger} \sigma^{j}&= -\sqrt{\frac{3}{2}} S^{ij \dagger} -\frac{1}{2} \iu \epsilon^{ijk} S^{k \dagger} \; , \nonumber \\
	\sigma^{i} S^{j} &= -\sqrt{\frac{3}{2}} S^{ij} -\frac{1}{2} \iu \epsilon^{ijk} S^{k} \; , \nonumber \\
	S^{i \dagger} S^{j} &= \frac{1}{3} \delta^{ij} \1 -\frac{1}{3} \Sigma^{ij} + \frac{1}{6} \iu \epsilon^{ijk} \Sigma^{k}\nonumber \; , \\
	S^{i} S^{j \dagger} &= \frac{1}{3}\bigl(2 \delta^{ij} \1- \iu \epsilon^{ijk} \sigma^{k} \bigr) \; , \nonumber \\
	\Sigma^{i} S^{j \dagger} &= -\sqrt{\frac{3}{2}} S^{ij \dagger} + \frac{5}{2} \iu \epsilon^{ijk} S^{k \dagger} \; , \nonumber \\
	S^{i} \Sigma^{j} &= -\sqrt{\frac{3}{2}} S^{ij} + \frac{5}{2} \iu \epsilon^{ijk} S^{k} \; ,  \nonumber \\
	\Sigma^{i}\Sigma^{j}&= 5 \delta^{ij} \1 + 4 \Sigma^{ij} +\iu \epsilon^{ijk} \Sigma^{k}   \; ,
	\label{eq:spin_matrices_relations}
	\end{align}
	which are verified by using their explicit representations in \cref{eq:sigma_matrices,eq:Spin_transition_matrices,eq:Sigma_matrices}.
	
\end{itemize}

\section{Imaginary parts of the loop functions arising from $2\pi$-exchange box diagrams}
\label{sec:coefficients_imaginary_parts}
\noindent
In this appendix, we list the imaginary parts of the loop functions introduced in \cref{sec:box} to specify the $2\pi$-exchange potentials from box diagrams.
We use the abbreviation $w= \sqrt{\mu^2 - 4 \mpi^2}$. 
\begin{align}
\Im A_2^{NN -} =& - \frac{w}{8 \mu \pi}  \nonumber\\
\Im A_3^{NN -} =& - \frac{w}{16 \mu \pi}   \nonumber\\
\Im A_4^{NN -} =& - \frac{w^3}{96 \mu \pi}   \nonumber\\
\Im B_2^{NN -} =& - \frac{\mu^2-2\mpi^2}{4 \mu^3 w \pi}    \nonumber\\
\Im B_3^{NN -} =& - \frac{\mu^2-3\mpi^2}{4 \mu^3 w \pi}   \nonumber\\
\Im B_4^{NN -} =& - \frac{\mu^4 -5 \mu^2 \mpi^2 +4\mpi^4}{24 \mu^3 w \pi}  \nonumber\\
\Im C_4^{NN -} =& - \frac{\mu^4 -4 \mu^2 \mpi^2 +2\mpi^4}{4 \mu^5 w \pi}  \\\nonumber\\
\Im A_2^{N\Delta -} =&  \frac{1}{32 \Delta \mu \pi} \biggl[  -2 \Delta  w - \frac{\pi}{2} w^2 + (4\Delta^2 +w^2) \arctan \frac{ w}{2 \Delta}  \biggr]  \nonumber\\
\Im A_3^{N\Delta -} =&  \frac{1}{64 \Delta \mu \pi} \biggl[  -2 \Delta  w - \frac{\pi}{2} w^2 + (4\Delta^2 +w^2) \arctan \frac{ w}{2 \Delta}  \biggr]  \nonumber\\
\Im A_4^{N\Delta -} =&  \frac{1}{512 \Delta \mu \pi} \biggl[  -8 \Delta^3 w - \frac{10}{3} \Delta  w^3 - \frac{\pi}{2} w^4 + (4\Delta^2 +w^2)^2 \arctan \frac{ w}{2 \Delta}  \biggr]  \nonumber\\
\Im B_2^{N\Delta -} =&  \frac{1}{32 \Delta \mu^3 \pi} \biggl[ - 2 \Delta w + \frac{\pi}{2} (4 \mpi^2-3\mu^2) + ( 4 \Delta^2-4 \mpi^2 + 3\mu^2)  \arctan \frac{ w}{2 \Delta}  \biggr]  \nonumber\\
\Im B_3^{N\Delta -} =&  \frac{1}{64 \Delta \mu^3 \pi} \biggl[ - 6 \Delta w + \frac{\pi}{2} (12 \mpi^2 - 5 \mu^2) + (12\Delta^2 -12 \mpi^2 + 5 \mu^2)  \arctan \frac{ w}{2 \Delta}  \biggr]  \nonumber\\
\Im B_4^{N\Delta -} =&  \frac{1}{512 \Delta \mu^3 \pi} \biggl[ -\frac{2}{3} \Delta w (12 \Delta^2 - 20 \mpi^2 + 17 \mu^2) - \frac{\pi}{2} ( 16 \mpi^4 -24 \mpi \mu + 5\mu^4 )  \nonumber\\*
&+(4\Delta^2 - 4 \mpi^2 + \mu^2)(4\Delta^2 - 4 \mpi^2 + 5\mu^2)  \arctan \frac{ w}{2 \Delta}  \biggr]  \nonumber\\
\Im C_4^{N\Delta -} =&  \frac{1}{512 \Delta \mu^5 \pi} \biggl[ -2 \Delta w( 12 \Delta^2 - 20 \mpi^2 + 29 \mu^2 )  - \frac{\pi}{2} ( 48 \mpi^4 - 120 \mpi^2 \mu^2 +35 \mu^2) \nonumber\\*
&+ (48 (\Delta^2 -\mpi^2)^2 + 120 ( \Delta^2 - \mpi^2) \mu^2 + 35 \mu^2 ) \arctan \frac{ w}{2 \Delta}  \biggr]  \\\nonumber\\
\Im A_2^{\Delta\Delta -} =&  \frac{1}{64 \Delta \mu \pi} \biggl[  -6 \Delta  w - \frac{\pi}{2} w^2 + (12\Delta^2 +w^2) \arctan \frac{ w}{2 \Delta}  \biggr]  \nonumber \\
\Im A_3^{\Delta\Delta -} =&  \frac{1}{128 \Delta \mu \pi} \biggl[  -6 \Delta w - \frac{\pi}{2} w^2  + (12\Delta^2 +w^2) \arctan \frac{w}{2 \Delta}  \biggr]  \nonumber \\
\Im A_4^{\Delta\Delta -} =&  \frac{1}{1024 \Delta \mu \pi} \biggl[  - \frac{2}{3} \Delta w (60 \Delta^2 -52 \mpi^2 + 13\mu^2) - \frac{\pi}{2} w^4   \nonumber\\*
&+ (4\Delta^2 +w^2)(20\Delta^2 +w^2) \arctan \frac{w}{2 \Delta}  \biggr]  \nonumber\\*
\Im B_2^{\Delta\Delta -} =&  \frac{1}{64 \Delta \mu^3 \pi} \biggl[  -2 \Delta w  \frac{12 \Delta^2 -12 \mpi^2 + 5\mu^2}{w^2 + 4\Delta^2} + \frac{\pi}{2} (4\mpi^2-3\mu^2)   \nonumber\\*
&+ (12 \Delta^2 -4 \mpi^2 + 3\mu^2) \arctan \frac{w}{2 \Delta}  \biggr]  \nonumber \\
\Im B_3^{\Delta\Delta -} =&  \frac{1}{128 \Delta \mu^3 \pi} \biggl[  - 2 \Delta w \frac{36 \Delta^2 -36 \mpi^2 + 11\mu^2}{w^2 + 4\Delta^2} + \frac{\pi}{2} (12 \mpi^2-5\mu^2)  \nonumber\\*
&+ (36\Delta^2 -12 \mpi^2 + 5\mu^2) \arctan \frac{w}{2 \Delta}  \biggr]  \nonumber \\
\Im B_4^{\Delta\Delta -} =&  \frac{1}{1024 \Delta \mu^3 \pi} \biggl[  - \frac{2}{3} \Delta w (60 \Delta^2 -52 \mpi^2 +49 \mu^2) - \frac{\pi}{2} (16 \mpi^4 -24 \mpi^2 + 5 \mu^2)  \nonumber\\*
&+ (80 \Delta^4 - 96 \Delta^2 \mpi^2 + 72 \Delta^2 \mu^2 + 16 \mpi^4 - 24 \mpi^2 \mu^2 + 5 \mu^4) \arctan \frac{w}{2 \Delta}  \biggr]  \nonumber \\
\Im C_4^{\Delta\Delta -} =&  \frac{1}{1024 \Delta \mu^5 \pi} \biggl[  -2 \Delta w \Bigl(60 \Delta^2 - 52 \mpi^2 +85 \mu^2 + \frac{8 \mu^2}{w^2+ 4 \Delta^2}\Bigr) \nonumber\\*
& - \frac{\pi}{2} (48 \mpi^4 - 120 \mpi^2 \mu^2 + 35 \mu^2)  + (240 \Delta^4 - 288 \Delta^2 \mpi^2 + 360 \Delta^2 \mu^2 \nonumber\\*
&+ 48 \mpi^4 - 120 \mpi^2 \mu^2 + 35 \mu^4) \arctan \frac{w}{2 \Delta}  \biggr]   \\\nonumber\\
\Im A_2^{\Delta\Delta +} =& \frac{1}{64 \Delta \mu \pi} \biggl[  2 \Delta  w - \frac{\pi}{2} w^2 + (-4\Delta^2 +w^2) \arctan \frac{ w}{2 \Delta}  \biggr]  \nonumber\\
\Im A_3^{\Delta\Delta +} =& \frac{1}{128 \Delta \mu \pi} \biggl[  2 \Delta  w - \frac{\pi}{2} w^2 + (-4\Delta^2 +w^2) \arctan \frac{ w}{2 \Delta}  \biggr]    \nonumber\\
\Im A_4^{\Delta\Delta +} =& - \frac{1}{2048 \Delta \mu \pi} \biggl[ 4\Delta w (12 \Delta^2 + w^2)  - w^4 \pi  - 2 (12\Delta^2 -w^2)(4\Delta^2 +w^2) \arctan \frac{ w}{2 \Delta}  \biggr]  \nonumber\\
\Im B_2^{\Delta\Delta +} =& \frac{1}{128 \Delta \mu^3 \pi} \biggl[  4 \Delta  w \frac{4\Delta^2 -4 \mpi^2 +3 \mu^2}{w^2+ 4\Delta^2} +(4 \mpi^2 - 3 \mu^2) \pi \nonumber\\*
&- 2 (4 \Delta^2 + 4 \mpi^2 - 3 \mu^2) \arctan \frac{ w}{2 \Delta} \biggr] \nonumber\\
\Im B_3^{\Delta\Delta +} =& \frac{1}{256 \Delta \mu^3 \pi} \biggl[ 4 \Delta  w  \frac{12\Delta^2 -12 \mpi^2 +5 \mu^2}{w^2+ 4\Delta^2}  +(12 \mpi^2 - 5 \mu^2) \pi \nonumber\\*
&- 2 (12 \Delta^2 + 12 \mpi^2 - 5 \mu^2) \arctan \frac{ w}{2 \Delta} \biggr]  \nonumber\\
\Im B_4^{\Delta\Delta +} =& \frac{1}{1024 \Delta \mu^3 \pi} \biggl[  2 \Delta w (12 \Delta^2-4 \mpi^2+5 \mu^2)  +\frac{\pi}{2} (-16 \mpi^4+24 \mpi^2 \mu^2-5 \mu^4) \nonumber\\*
&+(-48 \Delta^4+32 \Delta^2 \mpi^2-24 \Delta^2 \mu^2+16 \mpi^4-24 \mpi^2 \mu^2+5 \mu^4 )\arctan \frac{ w}{2 \Delta} \biggr] \nonumber\\
\Im C_4^{\Delta\Delta +} =& \frac{1}{2048 \Delta \mu^5 \pi} \biggl[ 4 \Delta  w \Bigl( 4 \Delta^2+20 \mpi^2 +35 \mu^2 + \frac{128 (\Delta^2-\mpi^2)^2}{w^2+ 4\Delta^2}  \Bigr)   \nonumber\\*
& - (48 \mpi^4-120 \mpi^2 \mu^2+35 \mu^4) \pi  -2 (144 \Delta^4-96 \Delta^2 \mpi^2+120 \Delta^2 \mu^2  \nonumber\\*
&-48 \mpi^4+120 \mpi^2 \mu^2-35 \mu^4 )\arctan \frac{ w}{2 \Delta} \biggr] 
\end{align}

\section{Next-to-leading order contact potential including deltas}
\label{sec:delta_contact_nlo}
Using the definitions $\vec q = \vec p^{\;\prime}- \vec p$ and $\vec k = \frac{1}{2} ( \vec p^{\;\prime} + \vec p\;)$ for initial and final center-of-mass momenta $\vec p$ and $\vec p^{\;\prime}$, the contact potentials with external deltas for the coupled (NN, N$\Delta$, $\Delta$N, $\Delta\Delta$) channels at next-to-leading order read:
\begin{align}
V_{ct,NNN\Delta}^{(2)} =&  \left(C_{3,NNN\Delta}^{(2)} q^2 + C_{4,NNN\Delta}^{(2)} k^2\right)  \; \vec \sigma_1 \cdot \vec S_2^{\dagger} + C_{5,NNN\Delta}^{(2)} \iu \vec  S_2^{\dagger} \cdot (\vec{q}\times \vec{k}) \nonumber\\*
&+ C_{6,NNN\Delta}^{(2)} \; \vec \sigma_1 \cdot \vec{q} \,\, \vec S_2^{\dagger} \cdot \vec{q}
+ C_{7,NNN\Delta}^{(2)} \; \vec \sigma_1 \cdot \vec{k} \,\, \vec S_2^{\dagger} \cdot \vec{k} \nonumber\\
V_{ct,N\Delta N\Delta}^{(2)} =& C_{1,N\Delta N\Delta}^{(2)} q^2 + C_{2,N\Delta N\Delta}^{(2)} k^2 
+ \left(C_{3,N\Delta N\Delta}^{(2)} q^2 + C_{4,N\Delta N\Delta}^{(2)} k^2\right)  \; \vec \sigma_1 \cdot \vec \Sigma_2 \nonumber\\*
&+ C_{5,N\Delta N\Delta}^{(2)} \iu ( \vec \sigma_{1} + \vec \Sigma_{2} ) \cdot (\vec{q}\times \vec{k})
+ C_{5-,N\Delta N\Delta}^{(2)} \iu ( \vec \sigma_{1} - \vec \Sigma_{2} ) \cdot (\vec{q}\times \vec{k})\nonumber\\*
&+ C_{6,N\Delta N\Delta}^{(2)} \; \vec \sigma_1 \cdot \vec{q} \,\, \vec \Sigma_2 \cdot \vec{q}
+ C_{7,N\Delta N\Delta}^{(2)} \; \vec \sigma_1 \cdot \vec{k} \,\, \vec \Sigma_2 \cdot \vec{k} \nonumber\\
V_{ct,NN\Delta \Delta}^{(2)} =& \left(C_{3,N N \Delta \Delta}^{(2)} q^2 + C_{4,N N \Delta \Delta}^{(2)} k^2\right)  \; \vec S_1^{\dagger} \cdot \vec S_2^{\dagger} \nonumber\\*
&+ C_{6,N N \Delta \Delta}^{(2)} \; \vec S_1^{\dagger} \cdot \vec{q} \,\, \vec S_2^{\dagger} \cdot \vec{q}
+ C_{7,N N \Delta \Delta}^{(2)} \; \vec S_1^{\dagger} \cdot \vec{k} \,\, \vec S_2^{\dagger} \cdot \vec{k} \nonumber\\*
&+ \left(C_{8,N N \Delta \Delta}^{(2)} q^2 + C_{9,N N \Delta \Delta}^{(2)} k^2\right)  \; S_1^{ij\dagger}  S_2^{ij\dagger} \nonumber\\*
&+ \left(C_{10,N N \Delta \Delta}^{(2)} q^i q^k + C_{11,N N \Delta \Delta}^{(2)} k^i k^k\right)  \; S_1^{ij\dagger}  S_2^{kj\dagger} \nonumber\\
V_{ct,N\Delta \Delta \Delta}^{(2)} =& \left(C_{3,N \Delta \Delta \Delta}^{(2)} q^2 + C_{4,N \Delta \Delta \Delta}^{(2)} k^2\right)  \; \vec S_1^{\dagger} \cdot \vec \Sigma_2  + C_{5,N\Delta\Delta\Delta}^{(2)} \iu  \vec  S_1^{\dagger} \cdot (\vec{q}\times \vec{k}) \nonumber\\*
&+ C_{6,N \Delta \Delta \Delta}^{(2)} \; \vec S_1^{\dagger} \cdot \vec{q} \,\, \vec \Sigma_2 \cdot \vec{q}
+ C_{7,N \Delta \Delta \Delta}^{(2)} \; \vec S_1^{\dagger} \cdot \vec{k} \,\, \vec \Sigma_2 \cdot \vec{k} \nonumber\\*
&+ \left(C_{8,N \Delta \Delta \Delta}^{(2)} q^2 + C_{9,N \Delta \Delta \Delta}^{(2)} k^2\right)  \; S_1^{ij\dagger}  \Sigma_2^{ij} \nonumber\\*
&+ \left(C_{10,N \Delta \Delta \Delta}^{(2)} q^i q^k + C_{11,N \Delta \Delta \Delta}^{(2)} k^i k^k\right)  \; S_1^{ij\dagger}  \Sigma_2^{kj} \nonumber\\
V_{ct,\Delta \Delta \Delta \Delta}^{(2)} =& C_{1,\Delta \Delta \Delta \Delta}^{(2)} q^2 + C_{2,\Delta \Delta \Delta \Delta}^{(2)} k^2 
+ \left(C_{3,\Delta \Delta \Delta \Delta}^{(2)} q^2 + C_{4,\Delta \Delta \Delta \Delta}^{(2)} k^2\right)  \; \vec \Sigma_1 \cdot \vec \Sigma_2 \nonumber\\*
&+ C_{5,\Delta \Delta \Delta \Delta}^{(2)} \iu ( \vec \Sigma_{1} + \vec \Sigma_{2} ) \cdot (\vec{q}\times \vec{k})\nonumber\\*
&+ C_{6,\Delta \Delta \Delta \Delta}^{(2)} \; \vec \Sigma_1 \cdot \vec{q} \,\, \vec \Sigma_2 \cdot \vec{q}
+ C_{7,\Delta \Delta \Delta \Delta}^{(2)} \; \vec \Sigma_1 \cdot \vec{k} \,\, \vec \Sigma_2 \cdot \vec{k} \nonumber\\*
&+ \left(C_{8,\Delta \Delta \Delta \Delta}^{(2)} q^2 + C_{9,\Delta \Delta \Delta \Delta}^{(2)} k^2\right)  \; \Sigma_1^{ij}  \Sigma_2^{ij} \nonumber\\*
&+ \left(C_{10,\Delta \Delta \Delta \Delta}^{(2)} q^i q^k + C_{11,\Delta \Delta \Delta \Delta}^{(2)} k^i k^k\right)  \; \Sigma_1^{ij}  \Sigma_2^{kj} \nonumber\\*
&+ \left(C_{12,\Delta \Delta \Delta \Delta}^{(2)} q^2 + C_{13,\Delta \Delta \Delta \Delta}^{(2)} k^2\right)  \; \Sigma_1^{ijk}  \Sigma_2^{ijk} \nonumber\\*
&+ \left(C_{14,\Delta \Delta \Delta \Delta}^{(2)} q^i q^l + C_{15,\Delta \Delta \Delta \Delta}^{(2)} k^i k^l\right)  \; \Sigma_1^{ijk}  \Sigma_2^{ljk}
\label{eq:contact_potential_nlo} \;,
\end{align}
where the purely nucleonic part $V_{ct,NNNN}^{(2)}$ is written in \cref{eq:contact_potential_nlo_NNNN}.
\section{Formulas for deuteron properties in the presence of $\Delta\Delta$-components}
When including the $\Delta\Delta$-channel with total isospin $I=0$, the
wave function of the deuteron gets extended and it involves six radial
functions. For the NN-state one has the usual assignment between partial
waves and radial functions: $\pwave{3S1}\to u(r)$ and $\pwave{3D1}\to w(r)$,
while the $\Delta\Delta$-state introduces four additional (small)
components: $\pwave{3S1}\to \widetilde u(r)$, $\pwave{3D1}\to \widetilde
w(r)$,
$\pwave{7D1}\to \tilde w_7(r)$, and $\pwave{7G1}\to \tilde v(r)$.

The normalization condition reads now:
\begin{align} \int_0^\infty \!\,dr\big[u(r)^2+w(r)^2+\widetilde u(r)^2+
  \widetilde w(r)^2+\widetilde w_7(r)^2+\widetilde v(r)^2\big] =1\,,
\end{align}
and the matter radius $r_d$ of the deuteron is calculated as:
\begin{align} r_d={1\over 2} \Big\{\int_0^\infty \!dr\,r^2\big[u(r)^2+w(r)^2
 +\widetilde u(r)^2+ \widetilde w(r)^2+\widetilde w_7(r)^2+\widetilde
  v(r)^2\big]\Big\}^{1/2}\,.\end{align}
The extended formula for computing the deuteron quadrupole moment $Q_d$ is
given by:
\begin{align} Q_d =& {1\over 20} \int_0^\infty \!\,dr\,r^2 \Big\{w(r)\big[
\sqrt{8}\,u(r)-w(r)\big] \nonumber \\ &+\widetilde w(r)\big[ \sqrt{8}\,
\widetilde u(r)-\widetilde w(r)\big] + {2\over 7}\,\widetilde w_7(r)\big[
6\sqrt{3}\,\widetilde v(r)-  \widetilde w_7(r)\big] - {5\over 7}\,
\widetilde v(r)^2\Big\} \,.\end{align}
In the derivation one exploits the orthogonality of NN- and
$\Delta\Delta$-components, as well as the orthogonality of $S=1$ (triplet)
and $S=3$ (septet) spin wave functions.                                      


\end{document}